\documentclass[preprint,11pt]{elsarticle}

\usepackage{amsmath,amssymb,amsthm,amsfonts}
\usepackage{mathabx}
\usepackage{mathrsfs}
\usepackage{mathtools}
\usepackage{bm}
\usepackage{graphicx}
\usepackage[caption=false]{subfig}
\usepackage{color,xcolor}	
\usepackage{tabularx}
\biboptions{compress}	
\usepackage{epsfig,psfrag}
\usepackage{tikz}
\usepackage{pgflibraryshapes}
\usepackage{graphicx,psfrag}
\usepackage{hyperref}
\usepackage{float}
\usepackage{subfig}
\usepackage{changes}
\makeatletter
\@namedef{Changes@AuthorColor}{blue}
\colorlet{Changes@Color}{blue}
\makeatother

\setcitestyle{authoryear,round}

\makeatletter
\hypersetup{colorlinks=true}
\AtBeginDocument{\@ifpackageloaded{hyperref}
	{\def\@linkcolor{magenta}
		\def\@anchorcolor{black}
		\def\@citecolor{teal}
		\def\@filecolor{cyan}
		\def\@urlcolor{magenta}
		\def\@menucolor{red}
		\def\@pagecolor{cyan}
		\begingroup
		\@makeother\`%
		\@makeother\=%
		\edef\x{%
			\edef\noexpand\x{%
				\endgroup
				\noexpand\toks@{%
					\catcode 96=\noexpand\the\catcode`\noexpand\`\relax
					\catcode 61=\noexpand\the\catcode`\noexpand\=\relax
				}%
			}%
			\noexpand\x
		}%
		\x
		\@makeother\`
		\@makeother\=
	}{}}
\makeatother

\makeatletter \oddsidemargin 0.0cm \evensidemargin 0.0cm
\newcommand{\be}{\begin{equation}}
\newcommand{\en}{\end{equation}}

\def\bm#1{\mbox{\boldmath{$#1$}}}

\let\underbrace\LaTeXunderbrace

\numberwithin{equation}{section}
\theoremstyle{plain}

\newtheorem{theorem*}{Theorem}

\theoremstyle{definition}

\DeclareMathOperator{\tr}{tr}

\DeclareMathOperator{\sym}{sym}

\journal{Journal of the Mechanics and Physics of Solids}

\begin{document}

\begin{frontmatter}


\title{\textbf{An asymptotically consistent morphoelastic shell model for compressible biological structures with finite-strain deformations}}

\author[mymainaddress]{Xiang Yu}

\address[mymainaddress]{Department of Mathematics, School of Computer Science and Technology, Dongguan University of Technology, Dongguan, 523808, China}

\author[mysecondaryaddress]{Xiaoyi Chen \corref{mycorrespondingauthor}}
\cortext[mycorrespondingauthor]{Corresponding author}
\ead{ xiaoyichen@uic.edu.cn}

\address[mysecondaryaddress]{Guangdong Provincial Key Laboratory of Interdisciplinary Research and Application for Data Science, Department of Applied Mathematics, BNU-HKBU United International College, Zhuhai, 519087, China}

\begin{abstract}

We derive an asymptotically consistent morphoelastic shell model to describe the finite deformations of biological tissues using  the variational asymptotical method. Biological materials may exhibit remarkable compressibility when under  large deformations, and we take this factor into account for accurate predictions of their morphoelastic changes.  The morphoelastic shell model combines the growth model of Rodriguez {\it et al.} and a novel shell model developed by us. We start from the three-dimensional (3D) morphoelastic model and construct the optimal shell energy based on a series expansion around the middle surface. A two-step variational method is applied that retains the leading-order expansion coefficient while eliminating the higher-order ones.  The main outcome is a two-dimensional (2D) shell energy depending on the stretching and bending strains of the middle surface. The derived morphoelastic shell model is asymptotically consistent with three-dimensional morphoelasticity and can recover various shell models in literature. Several examples are shown for the verification and illustration.

\end{abstract}

\end{frontmatter}

\section{Introduction}

\begin{figure}[h!]
	\centering
	\includegraphics[width=0.9\linewidth]{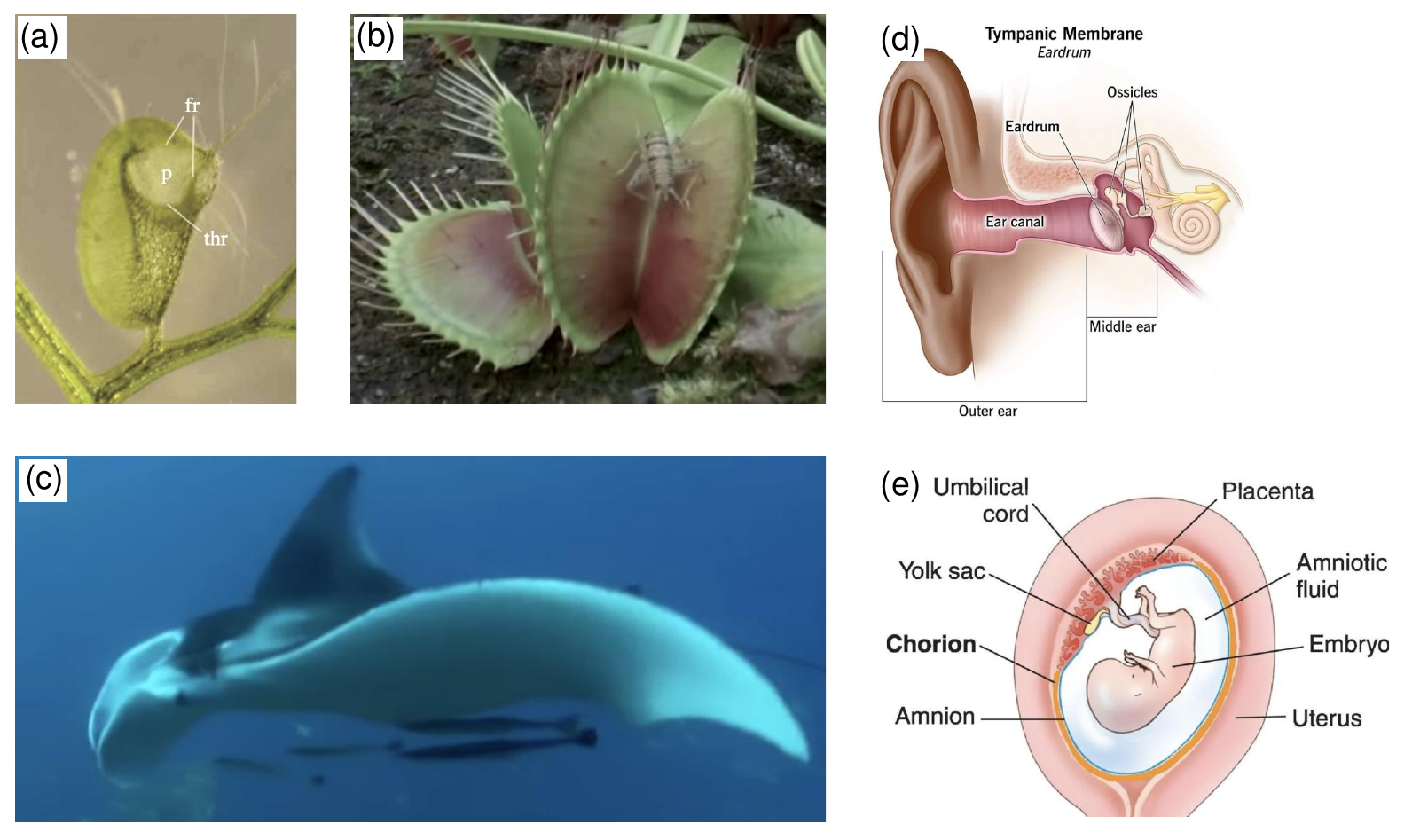}
	\caption{Examples of biological shells. (a) Leaves of the Utricularia (bladderworts). (b) Leaves of the Venus flytrap. (c) Pectoral fins of a manta ray. (d) Tympanic membrane (eardrum) of the human ear. (e) Fetal membrane (chorion and amnion) of the embryo. (a) is adapted from \cite{guo2015fast} under permission, (b)--(e) are taken from the websites.}
	\label{fig1}
\end{figure}

Many biological systems are shell-like structures with one dimension much smaller than the other two, for example,  the leaves of the Utricularia (bladderworts) and the  Venus flytrap, the triangular pectoral fins of manta rays, the tympanic membranes of our ears and fetal membranes of embryos; see Fig. \ref{fig1}. One feature of biological systems is that they grow to adapt to the environment, which is important for their normal functioning. As shown in Fig. \ref{fig1}, the bladderworts capture the prey by quickly expanding their leaves to generate a flow to suck the prey in; the Venus flytrap captures insects by fast changing the curvature of its leaves; the manta ray swims with the large bending of its pectoral fins; the tympanic membrane separates our inner and outer ears and transmits the sound from the air to our middle ear through finite deformations and the fetal membrane undergoes large deformations as the embryo is developing throughout the pregnancy period.  To better understand the mechanism under the large deformations of various thin biological shells, a morphoelastic shell theory with wide applicability needs to be developed.

By incorporating the growth effects,  \cite{dervaux2008morphogenesis} and \cite{dervaux2009morphogenesis} derived the generalized Föppl-von Kármán theory of thin plates  under the membrane assumption. As applications, they studied the morphogenesis of Acetabularia algae and grass blades  through linear bifurcation analysis. This morphoelastic  plate theory was also adopted by \cite{xu2020water} to study the patterns of lotus leaves, which can either be local wrinkles with short wavelength at the edge or the global bending cone with long rippled waves near the edge. The water as a substrate was shown to be the main contributor to this difference.  The Föppl--von Kármán plate theory of growth was later improved by Wang and coworkers \citep{wang2018consistent,li2022analytical,du2023general,du2023simplified}  to account for  finite-strain deformations.  They started from a series expansion of the displacement without {\it ad hoc} assumptions. Then through asymptotic ally consistent manipulation of the three-dimensional governing system, plate equations were derived. The boundary conditions were proposed separately. Following similar ideas, \cite{li2023general} constructed a shell theory incorporating the growth effects for incompressible materials. Another morphoelastic shell model was established by \cite{haas2021morphoelasticity} to account for the large bending of the incompressible cell sheet. Proper scaling for the intrinsic curvature is introduced and asymptotic expansion of the energy is implemented to derive the shell theory in the limit of thin shells. 
 \cite{yin2022three} developed a chemomechanical framework for active shells, incorporating both biochemical and mechanical factors, which successfully accounts for the mechanical feedback on cellular chemical patterns. Their framework employs Koiter's shell theory, suitable for analyzing deformations with large rotations but small strains.

Apart from the continuum models, some discrete computational models are also obtained to describe the mechanics of the growing thin membranes and shells.  \cite{rausch2014mechanics} derived a finite element model to describe the growing biological membranes using discrete Kirchhoff shell kinematics. By applying this model, they demonstrated the chronic adaptation of the leaflet membrane under pathological loading conditions.  \cite{rudraraju2019computational} constructed a three-dimensional computational framework coupling morphology, incremental surface growth by accretion and morphoelastic volume growth to study the shape evolution of mollusk  shells.

The biological shells may also undergo stress-free deformations with rich morphologies. Chen and  coworkers \citep{chen2020stress,chen2021physical,chen2022generating}  developed a stress-free growth framework by appropriately  choosing  the forms of the growth tensor and the elastic deformation tensor, which allows analytical expressions of the growth functions under special conditions. The obtained growth formulas are able to capture various morphogenesis of the biological tissues, including the  fern uncurling, the mushroom cap expansion, and the pattern formation of the brain organoid and the bitter gourd with two rows of singular cusps.
 \cite{dai2022minimizing} further developed  the stress-free growth framework 
through conformal mappings and represented the shape of the leaves through holomorphic functions, which can capture the convex, concave or sharp-pointed geometries. This flexible method is used to study the fenestration process in Monstera deliciosa.

The aforementioned morphoelastic models use different approximate approaches to treat the shell deformations, including  the classical plate/shell theories built on asymptotic derivations, improved plate/shell theories based on series expansions, and discrete numerical models. However, from a mathematical point of view,  there still exist  some unsatisfactory features. For example, classical models like the Kirchhoff--Love and F\"{o}ppl--von K\'{a}m\'{a}n models  are suitable for small strain problems and are established on a set of prior assumptions, which limit their applicability for finite-strain deformations. Most series-expansion-based approaches   (e.g., \cite{dai2014consistent,wang2018consistent,mehta2022wrinkling}) start from the local differential equations and thus lose the variational structures of the original problem, which may cause certain inconsistency in numerical calculations.
\cite{steigmann2008two,steigmann2012extension,steigmann2015mechanics} aimed to resolve this issue by employing a mixed approach---using both the energy functional and governing equations.
However, the sacrifice is the loading condition---the top and bottom surfaces are subjected to nearly traction-free conditions.  We note a recent work by  \cite{carrera2022carrera}, which uses the series expansion and virtual displacement method for the derivation.

\begin{figure}[h!]
	\centering
	\includegraphics[width=0.8\linewidth]{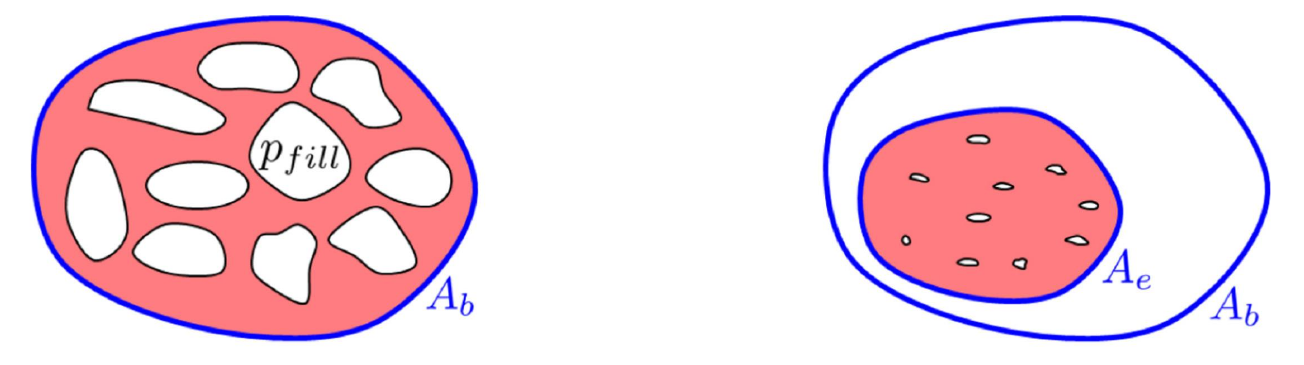}
	\caption{Demonstration of the lung parenchymal during the gas exchange process, showing the compressibility of biological tissues. Adapted from \cite{birzle2018experimental} under permission.}
	\label{fig2}
\end{figure}

We also want to mention that when studying the large deformation of biological tissues, the compressibility of biological materials can be an important factor. Although biological materials are normally modeled as incompressible, they  may exhibit remarkable compressibility when under large deformations. A good example demonstrating this is  the deformation of the lung parenchyma in the gas exchange process, during which large volume changes are observed; see Fig. \ref{fig2}. 
The renowned study by \cite{chuong1984compressibility} demonstrated slight compressibility ($0.5\%$--$1.26\%$) in the thoracic arteries of four rabbits under a radial compressive stress load of $10$ KPa. 
 Many researchers cited this paper as evidence to support the incompressibility of the arteries or biological tissues since the relative volume variation is indeed very small. 
The assumption for the incompressibility of biological tissues is reasonable under certain conditions when the strains are small, just as that in \cite{chuong1984compressibility}. This may be because, under the small strain, the tissues are still able to hold the fluid. Besides, the incompressibility condition can greatly simplify the constitutive modeling as well as the computation, which also motivates the researchers to use this condition.  However, experimental evidence gas shown that some biological tissues do exhibit remarkable compressibility even within physiological conditions. \cite{tickner1967theory} showed that the relative volume change of the brachial artery of humans can reach as high as $35\%$, and  \cite{chesler2004measurements}  found that the left pulmonary artery of the mouse can have compressibility of  $15\%$--$20\%$. Readers may refer to more recent works \citep{di2012review,nolan2016compressibility,yossef2017further} to learn more about the compressibility of biological tissues in experiments. To provide more accurate predictions for biological tissues under finite-strain deformations, we consider the factor of compressibility in our model.

In this paper, we derive a consistent finite-strain morphoelastic shell model that takes the compressibility into account using the variational asymptotic method pioneered. This method, which was first initiated by \cite{berdichevskii1979variational}, is an combination of variational principles and asymptotic approaches: variational principles are used to defined functionals and asymptotic approaches are applied to the same functional instead of applying on differential equation which is more prone to error. It has been applied to successfully  deliver highly accurate reduced model for a wide range of elasticity problems, including necking in prismatic solids \citep{audoly2016analysis}, elastocapillary necking of cylindrical gels, bulging of rubber tubes, and morphoelastic rods; see \cite{lestringant2020asymptotically} for a summary. Here, starting from three-dimensional energy functional  of a morphoelastic shell, we first construct an approximate energy functional that involves higher-order expansion coefficients based on a series expansion of the displacement around the middle surface. By fixing the leading-order coefficient and optimizing the energy functional with respect to the higher-order ones, we obtain the expressions of the latter in terms of the former by solving Euler-Lagrange equations. This results in a two-dimensional shell energy that depends only on the leading-order coefficient. To further derive the shell equations and boundary conditions, the shell energy is optimized again with respect to the leading-order coefficient. This two-step optimization scheme leads to a morphoelastic shell model that  has the following three key features:
\begin{enumerate}
	\item  Both geometric and material nonlinearities are retained in the shell model, which makes it suitable for studying finite-strain deformations of biological structures. 
	\item  The model captures both the stretching and bending effects in an asymptotically correct way, and can be applied under general loading conditions.
	\item  The reduced energy retains the variational structure of the three-dimensional morphoelasticity, which  can be readily applied in numerical simulations.
\end{enumerate}

The rest of this paper is arranged as follows. In Section \ref{sec:geometry}, we present the three-dimensional morphoelastic shell model, which is the full model for the dimension reduction. In Section \ref{sec:shell}, we derive the two-dimensional morphoelastic shell model, including the reduced energy as well as the shell equations and boundary conditions, 
which is the key part of this study. In Section \ref{sec:connection}, the classical shell models are recovered from our model. In Section \ref{sec:application}, we validate the derived shell model by two benchmark problems with  exact solutions. In Section \ref{sec:example}, the model is applied to study the invagination of the Volvox globator which is a spherical monolayer cell sheet (modeled as a thin biological shell). 
Finally, concluding remarks are given in Section \ref{sec:con}.

The mathematical notations used in this paper are as follows. Boldface letters, for example $\bm{x}$, $\bm{F}$, represent vectors or second-order tensors; calligraphy letters, like $\mathcal{A}$, $\mathcal{B}$, refer to higher-order tensors. The summation convention for repeated indices is adopted, in which  Greek letters  $\alpha,\beta,\dots$ run from $1$ to $2$, whereas Latin letters $i,j,k,\dots$ run from $1$ to $3$. A comma preceding indices means differentiation.  Let $\{\bm{e}_1,\bm{e}_2,\bm{e}_3\}$ be an orthonormal basis of the three-dimensional Euclidean space. The components of a tensor $\mathcal{R}$ of order $n$ with respect to this basis is written as $\mathcal{R}_{i_1\dots i_n}$. Given two tensors $\mathcal{A}$ and $\mathcal{B}$ be two tensors of order $p$ and $q$, respectively,   their simple and double  dot products (contractions) are defined as
\begin{align}
&(\mathcal{A}\cdot \mathcal{B})_{i_1\dots i_{p-1}j_2\dots j_q}= \mathcal{A}_{i_1\dots i_{p-1}k}\mathcal{B}_{kj_2\dots j_q},\\
&(\mathcal{A}:\mathcal{B})_{i_1\dots i_{p-2}j_3\dots j_q} = \mathcal{A}_{i_1\dots i_{p-2}kl}\mathcal{B}_{klj_3\dots j_q}.
\end{align}
In particular, the double dot product of two second-order tensors is given by $\bm{A}:\bm{B}=\tr(\bm{A}\bm{B}^T)$. Given a tensor-valued function $\mathcal{R}(\mathcal{A})$ of a tensor variable $\mathcal{A}$, where $\mathcal{R}(\mathcal{A})$ and $\mathcal{A}$ are tensors of order $p$ and $q$, respectively, we define its derivative as
\begin{align}
\Big(\frac{\partial \mathcal{R}}{\partial\mathcal{A}}\Big)_{i_1\dots i_pj_1\dots j_q}=\frac{\partial \mathcal{R}_{i_1\cdots i_p}}{\partial \mathcal{A}_{j_1\dots j_q}}.
\end{align}
Higher-order derivatives are defined in a similar way.

\section{Three-dimensional morphoelastic model}\label{sec:geometry}
The three-dimensional morphoelastic problem is introduced in this section, and for easier reading,  some knowledge of shell geometries is first reviewed.

\subsection{Geometry of shells}

\begin{figure}[h!]
	\centering
	\includegraphics[width=0.8\linewidth]{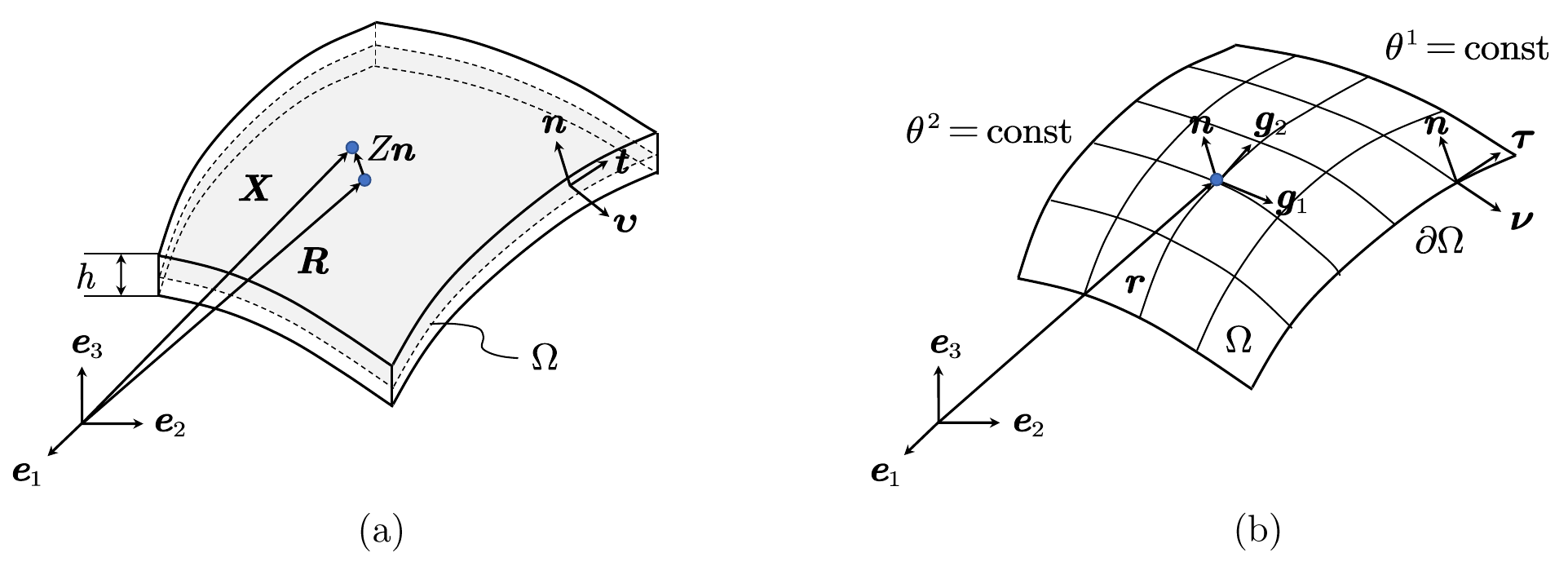}
	\caption{(a) A thin shell of thickness $h$ with a material point $\bm{R}$ on the middle surface $\Omega$ and a material point $\bm{X}$ in the shell body. (b) The curvilinear coordinates on the middle surface of the shell and the associated covariant basis $\{\bm{g}_1, \bm{g}_2, \bm{n}\}$. The covariant basis on the lateral surface are  $\{\bm{t}, \bm{\upsilon}, \bm{n} \}$ (see (a)) which are denoted as $\{\bm{\tau}, \bm{\nu}, \bm{n} \}$ on the middle surface.}
	\label{fig:shell}
\end{figure}

We consider a thin shell of constant thickness $h$ that occupies a region $\Omega\times [-h/2,h/2]$ in its reference configuration, where $\Omega$ is the middle surface of the shell with its boundary denoted by $\partial\Omega$; see Fig. \ref{fig:shell}(a). By a thin shell we mean that thickness is small  compared with the length scale of  other dimensions as well as the radius of curvature.  
Following the common practice \citep{ciarlet2005introduction, steigmann2012extension}, we parameterize the middle surface $\Omega$ locally by two curvilinear coordinates $\theta^{1}$ and $\theta^2$, as shown in Fig. \ref{fig:shell}(b). The position  of a generic point on $\Omega$ is written as $\bm{R}=\bm{R}(\theta^{1},\theta^2)$. The tangent vectors along the coordinate lines are given by $\bm{g}_\alpha={\partial\bm{R}}/{\partial\theta^\alpha}=\bm{R}_{,\alpha}$.
 Their contravariant counterparts are denoted by $\bm{g}^\alpha$, which satisfy the relation $\bm{g}^\alpha\cdot\bm{g}_\beta=\delta^\alpha_\beta$, where $\delta^\alpha_\beta$ is the Kronecker delta. The unit normal  to the middle surface is defined by $\bm{n}={\bm{g}_1\times\bm{g}_2}/{\sqrt{g}}$, where $\times$ means cross-product and $g=\det(\bm{g}_\alpha\cdot\bm{g}_\beta)$ is the metric determinant with $(\bm{g}_\alpha\cdot\bm{g}_\beta)$ denoting the $2\times 2$ matrix whose $\alpha\beta$-component is $\bm{g}_\alpha\cdot\bm{g}_\beta$. We set $\bm{g}_3=\bm{g}^3=\bm{n}$ so that $\{\bm{g}_i\}$ and $\{\bm{g}^i\}$ form two sets of right-handed bases. With the aid of the unit normal, the contravariant tangent vectors can also be expressed as $\bm{g}^\alpha=\frac{e^{\alpha\beta}}{\sqrt{g}}\,\bm{g}_\beta\times\bm{n}$,
where $e^{\alpha\beta}=e_{\alpha\beta}$ is the unit alternator ($e^{12}=-e^{21}=1$, $e^{11}=e^{22}=0)$.

 The change of the unit normal is described by the curvature tensor, which is defined by
\begin{align}
\label{eq:curvature}
\bm{\kappa}=-\frac{\partial \bm{n}}{\partial\bm{R}}=-\bm{n}_{,\alpha}\otimes \bm{g}^\alpha.
\end{align}
An important fact is that $\bm{\kappa}$ is a symmetric tensor, that is, $\bm{\kappa}=\bm{\kappa}^T$ (\cite{do2016differential}, p. 142).
Associated to the curvature tensor, the mean and Gaussian curvatures are defined by $H=\tr(\bm{\kappa})/2$ and $K=\det(\bm{\kappa})$, respectively. By the well-known Cayley-Hamilton theorem, $\bm{\kappa}$ satisfies the characteristic equation
\begin{align}\label{eq:kappaCH}
\bm{\kappa}^2-2H\bm{\kappa}+K\bm{1}=0,
\end{align}
where $\bm{1}=\bm{I}-\bm{n}\otimes\bm{n}$ denotes the two-dimensional identity tensor on the tangent plane of the middle surface. An immediate consequence of \eqref{eq:kappaCH} is that the cofactor $\bm{\kappa}^*:=\det(\bm{\kappa})\bm{\kappa}^{-T}$ can be expressed as 
\begin{align}\label{eq:kappastar}
\bm{\kappa}^*=K\bm{\kappa}^{-1}=2H\bm{1}-\bm{\kappa},
\end{align}
where we have used the symmetry of $\bm{\kappa}$.

For a shell structure, the position of a material point in the reference configuration is decomposed as
\begin{align}\label{eq:X}
\bm{X}=\bm{R}(\theta^1,\theta^2)+Z\bm{n}(\theta^1,\theta^2),\quad -\frac{h}{2}\leq Z\leq \frac{h}{2},
\end{align}
where $Z$ is the coordinate along the thickness direction.  The tangent vectors induced by  $\theta^\alpha$ at an arbitrary point inside the shell are denoted by $\tilde{\bm{g}}_\alpha$.  It follows from  \eqref{eq:X} that
\begin{align}\label{eq:ghat}
\tilde{\bm{g}}_\alpha=\frac{\partial\bm{X}}{\partial\theta^\alpha}=(\bm{1}-Z\bm{\kappa})\bm{g}_\alpha=\bm{\mu}\bm{g}_\alpha,
\end{align}
where  $\bm{\mu}$ is defined by
\begin{equation}
\label{Eq6_1}
\bm{\mu}=\bm{1}-Z\bm{\kappa}.
\end{equation}
Considering \eqref{eq:ghat}, one can calculate the volume element as
\begin{align}\label{eq:dV}
dV=(\tilde{\bm{g}}_1d\theta^1\times \tilde{\bm{g}}_2d\theta^2)\cdot \bm{n}dZ=\mu(Z)\,dAdZ,
\end{align}
where $\mu(Z)$ denotes the determinant of $\bm{\mu}$ and is given by
\begin{equation}
\label{eq:mu}
\mu(Z)=\det(\bm{\mu})=1-2HZ+KZ^2
\end{equation}
and $dA=(\bm{g}_1\times \bm{g}_2)\cdot\bm{n}\, d\theta^1d\theta^2$ represents the area element on the middle surface.

On the lateral surface $\partial\Omega\times[-h/2,h/2]$ of the shell, let $\bm{t}$ and $\bm{\upsilon}$ be the unit tangent and outward normal vectors, respectively, defined in a way such that $\bm{\upsilon}=\bm{t}\times\bm{n}$. Then on the middle surface, $\bm{t}$ and $\bm{\upsilon}$ are denoted as $\bm{\tau}$ and $\bm{\nu}$  respectively.
In view of \eqref{eq:ghat}, the vector area element on the lateral surface is calculated by
\begin{align}\label{eq:da}
\bm{\upsilon}\,da=\bm{\mu}\bm{\tau}dS\times \bm{n}dZ= (\bm{\mu}\bm{\tau}\times\bm{n})\,dZdS,
\end{align}
where $S$ denotes the arclength variable on $\partial\Omega$. With the use of the following equality (\cite{chadwick1999continuum}, p. 20)
\begin{align}
\bm{\mu}\bm{\tau}\times\bm{n}=\bm{\mu}^*(\bm{\tau}\times\bm{n}),
\end{align}
where $\bm{\mu}^*=\det(\bm{\mu})\bm{\mu}^{-T}$ is the cofactor of $\bm{\mu}$, we deduce from \eqref{eq:da} that
\begin{align}\label{eq:daa}
da=|\bm{\mu}^*\bm{\nu}|\,dZdS=\mu^*(Z)\,dZdS,
\end{align} 
where the last equation serves to define $\mu^*(Z)$. From \eqref{eq:kappastar} and \eqref{Eq6_1}, we have that 
\begin{align}
\bm{\mu}^*=\bm{1}-Z\bm{\kappa}^* =\bm{1}+Z(\bm{\kappa}-2H\bm{1})
\end{align}
and consequently
\begin{align}
\mu^*(Z)=|\bm{\mu}^*\bm{\nu}|=\sqrt{(1-Z\kappa_\tau)^2+(Zc_\tau)^2},
\end{align}
where we have used the decomposition $\bm{\kappa}=\kappa_\tau\bm{\tau}\otimes\bm{\tau}+\kappa_\nu\bm{\nu}\otimes\bm{\nu}+c_\tau(\bm{\tau}\otimes\bm{\nu}+\bm{\nu}\otimes\bm{\tau})$  on $\partial\Omega$.

\subsection{Variational formulation in three dimensions}

The shell  deforms under the combined loading of  biological growth and external forces or constraints.
  After deformation, the material point $\bm{X}=\bm{R}+Z\bm{n}$ moves to a new position written as $\bm{x}=\bm{x}(\bm{R},Z)$. The deformation gradient is obtained as
\begin{align}\label{eq:F}
\bm{F}=\frac{\partial \bm{x}}{\partial\bm{X}}=\bm{x}_{,\alpha}\otimes \tilde{\bm{g}}^\alpha+\bm{x}_{,Z}\otimes \bm{n}=(\nabla\bm{x})\bm{\mu}^{-1}+\bm{x}_{,Z}\otimes \bm{n},
\end{align}
where $\nabla=\bm{g}^\alpha\frac{\partial}{\partial \theta^\alpha}$ is the two-dimensional del operator on the middle surface  $\Omega$ and $\nabla\bm{x}=\bm{x}_{,\alpha}\otimes\bm{g}^\alpha$ denotes the surface gradient of $\bm{x}$. According to the basic assumption of growth theory by \cite{rodriguez1994stress}, the total deformation gradient $\bm{F}$ can be decomposed as follows:
\begin{align}\label{eq:FAG}
\bm{F}=\bm{A}\bm{G},
\end{align}
where  $\bm{G}$ is the growth tensor and $\bm{A}$ is the elastic deformation tensor.  To consider the compressibility of biological materials,  the elastic deformation  is not assumed to be isochoric, i.e., we do not use the condition $\det(\bm{A})=1$. This retains the generality of the theory, and the incompressible deformation can be regarded as the limiting case of the compressible one (e.g., letting Poisson's ratio go to $1/2$).

The biological shell is assumed to be hyperelastic with a strain energy density $W=W(\bm{A})$ per unit volume of the grown state, which satisfies the strong-ellipticity condition pointwise:
\begin{align}
\bm{a}\otimes \bm{b}:\frac{\partial^2 W}{\partial \bm{A}^2}:\bm{a}\otimes \bm{b}>0,\quad \forall\, \bm{a}\otimes \bm{b}\neq 0.
\end{align}
To deal with growth-induced deformations, it is convenient to introduce the following {\it augmented energy density function} \citep{ciarletta2012growth}
	\begin{align}\label{eq:Psi}
	\bar{W}(\bm{F})=\det(\bm{G}) W(\bm{F}\bm{G}^{-1}),
	\end{align}
where we have suppressed in notation the dependence of $\bar{W}$ on $\bm{G}$ as it plays on role in the subsequent derivation. The associated first Piola--Kirchhoff (P--K) stress can be written as $\bm{P}=\frac{\partial\bar{W}}{\partial\bm{F}}=\det(\bm{G})\frac{\partial W}{\partial \bm{A}}\bm{G}^{-T}$.

The shell is subjected to applied traction $\bm{q}^+$ and $\bm{q}^-$ on the top and bottom surfaces, respectively. On the lateral surface, prescribed traction $\bm{q}$ and position are imposed on   $\partial \Omega_u\times [-h/2,h/2]$ and $\partial\Omega_t\times[-h/2,h/2]$, respectively, where $\partial \Omega=\partial\Omega_u \cup \partial\Omega_t$. With the body force neglected, the total potential energy functional  of the shell is composed of the elastic energy and the load potential, which reads
\begin{align}\label{eq:E}
\begin{split}
\mathcal{E}[\bm{x}]=&\int_{\Omega}\int_{-h/2}^{h/2} \bar{W}(\bm{F})\mu(Z)\, dZdA-\int_{\Omega}\big(\bm{q}^{+}(\bm{R})\cdot\bm{x}(\bm{R},h/2) \mu(h/2)\\
&+\bm{q}^{-}(\bm{R})\cdot\bm{x}(\bm{R},-h/2)\mu(-h/2)\big)\,dA-\int_{\partial\Omega_t}\int_{-h/2}^{h/2}\bm{q}(S,Z)\cdot\bm{x}(S,Z)\mu^*(Z)\,dZdS,
\end{split}
\end{align}
The three-dimensional energy functional \eqref{eq:E} will be used as a starting point for the derivation of our morphoelastic shell model. 
\section{Two-dimensional morphoelastic shell model}\label{sec:shell}

In this section, we derive an asymptotically consistent two-dimensional morphoelastic shell model from the three-dimensional model  formulated in Section \ref{sec:geometry}. The asymptotic derivation builds on two main ingredients: one is  an asymptotic expansion method, and the other is a two-step variational method.

\subsection{Expansion of the energy functional}
Our first step is to  expand the energy function in terms of the thickness. For the purpose of dimension reduction, 
The position of a material point in the current configuration is expressed as the sum of a base surface's position (averaged position) and the material point's relative displacement from this base surface, i.e., 
\begin{align}\label{eq:xry}
\bm{x}(\bm{R},Z)=\bm{r}(\bm{R})+\bm{y}(\bm{R},Z).
\end{align}
We construct the {\it actual middle surface} as our base surface, which is defined as
\begin{align}\label{eq:rR1}
\bm{r}(\bm{R})=\frac{1}{h}\int_{-h/2}^{h/2}\bm{x}(\bm{R},Z)\,dZ.
\end{align}
It is a surface in the current configuration that passes through the centroid of the material line that is initially normal to the middle surface. Note that this surface typically differs from the deformed middle surface  $\bm{x}(\bm{R},0)$. The introduction of $\bm{r}(\bm{R})$ makes the dimension reduction process easier and more consistent, and is more applicable under complex loading conditions.

From (\ref{eq:rR1}) and (\ref{eq:xry}), we see that $\bm{y}$ satisfies the kinematic constraint
\begin{align}\label{eq:ycon}
\int_{-h/2}^{h/2}\bm{y}(\bm{R},Z)\,dZ=0.
\end{align}
Assume that all quantities involved are sufficiently smooth, then all variables can be expanded into Taylor series with respect to $Z$. In particular, the displacement $\bm{y}$ of a material point relative to the actual middle surface can be expanded as
\begin{equation}
\begin{aligned}\label{eq:y}
\bm{y}(\bm{R},Z)&=\bm{y}_1(\bm{R})Z+\frac{1}{2}\bm{y}_2(\bm{R})\Big(Z^2-\frac{1}{12}h^2\Big)+\frac{1}{6}\bm{y}_3(\bm{R})Z^3+O(h^4),\quad -\frac{h}{2}\leq Z\leq \frac{h}{2},
\end{aligned}
\end{equation}
where $\bm{y}_i=\frac{\partial^i \bm{y}}{\partial Z^i}|_{Z=0}, (i=1,2,3)$ are the expansion coefficients, and we also have from \eqref{eq:xry}  that   $\bm{y}_i=\frac{\partial^i \bm{x}}{\partial Z^i}|_{Z=0}, (i=1,2,3)$.  Note that we have used the kinematic constraint \eqref{eq:ycon} to eliminate the leading-order expansion coefficient of $\bm{y}$ to achieve the form of (\ref{eq:y}). Although \eqref{eq:y} looks like an ad hoc kinematic assumption, it can be derived from the asymptotic analysis  \citep{berdichevskii1979variational}, which is routine but slightly tedious, and justified rigorously by means of $\Gamma$-convergence \citep{friesecke2006hierarchy}. We use \eqref{eq:y} as a staring point to keep the derivation concise. The expansion coefficients $\bm{y}_1,\bm{y}_2,\bm{y}_3,\dots$ are called {\it directors} in the theory  of Cosserat continuum. The director $\bm{y}_1$ characterizes the direction of the material line initially normal to the middle surface, and the director $\bm{y}_2$ is related with the curvature of the same material line.

From \eqref{eq:F} as well as \eqref{eq:xry} and \eqref{eq:y}, the deformation gradient expanded to order $h^2$ is calculated as
\begin{align}\label{eq:Fa}
\bm{F}=\bm{F}_0+Z\bm{F}_1+\frac{1}{2}Z^2\bm{F}_2-\frac{1}{24} h^2\nabla\bm{y}_2,
\end{align}
where the coefficients are given by
\begin{align}
\begin{split}\label{eq:FF}
&\bm{F}_0=\nabla \bm{r}+\bm{y}_1\otimes \bm{n},\quad \bm{F}_1=(\nabla \bm{r})\bm{\kappa}+\nabla\bm{y}_1+\bm{y}_2\otimes\bm{n},\\
&\bm{F}_2=2 (\nabla\bm{r})\bm{\kappa}^2+2 (\nabla \bm{y}_1)\bm{\kappa}+\nabla\bm{y}_2+\bm{y}_3\otimes\bm{n}.
\end{split}
\end{align}
We see from \eqref{eq:Fa} that $\bm{F}_i=\frac{\partial^i \bm{F}}{\partial Z^i}|_{Z=0}$ for $i=1,2$ and $\bm{F}_0=\bm{F}|_{Z=0}+\frac{1}{24}h^2\nabla\bm{y}_2$. Thus in our nation, the subscript $i$ stands for the $i$-th derivative with respect to $Z$ evaluated at $Z=0$ for $i\geq 1$, while the subscript `$0$' does not only involves  the evaluation at $Z=0$, but also includes a correction term. 

To illustrate the dimension reduction process more clearly, we assume  in this section  that the growth tensor $\bm{G}$ is independent of $Z$. Nevertheless, one can easily adapt the derivation to the case that $\bm{G}$ is a function of $Z$, as what we have done for the examples in sections 5 and 6.. Given this assumption, expanding the strain energy $\bar{W}(\bm{F})$ at $\bm{F}=\bm{F}_0$ to order $h^2$  based on \eqref{eq:Fa} yields
\begin{align}\label{eq:W}
\bar{W}(\bm{F})=\bar{W}(\bm{F}_0)+Z \bm{P}_0:\bm{F}_1+\frac{1}{2}Z^2\bm{F}_1:\mathcal{A}:\bm{F}_1+\bm{P}_0:\frac{1}{2}\Big(Z^2\bm{F}_2-\frac{1}{12}h^2\nabla\bm{y}_2\Big),
\end{align}
where $\bm{P}_0=\frac{\partial\bar{W}}{\partial\bm{F}}({\bm{F}_0})$ and $\mathcal{A}=\frac{\partial^2\bar{W}}{\partial\bm{F}^2}({\bm{F}_0})$.

In the following, we will simplify the energy functional (\ref{eq:E}) based on the above expansions. 
Integrating (\ref{eq:W}) multiplied by $\mu(Z)$ from $Z=-h/2$ to $h/2$, we obtain
\begin{align}\label{eq:EE}
\begin{split}
\int_{-h/2}^{h/2} \bar{W}(\bm{F})\mu(Z)\,dZ=&h (1+\frac{1}{12} h^2 K)\bar{W}(\bm{F}_0)+\frac{1}{24}h^3\Big(\bm{F}_1:\mathcal{A}:\bm{F}_1+2\bm{P}_0:(\nabla\bm{r}\bm{\kappa}^2+ \nabla \bm{y}_1\bm{\kappa})\\
&+\bm{P}_0\bm{n}\cdot \bm{y}_3-4H \bm{P}_0:\bm{F}_1\Big)+O(h^4),
\end{split}
\end{align} 
where $\eqref{eq:FF}_3$ has been used to eliminate $\nabla\bm{y}_2$. 
The load potential per unit area due to the traction applied on the top and bottom surfaces is expanded as
\begin{align}
\begin{split}\label{eq:qtb}
&\bm{q}^+\cdot \bm{x}(\bm{R},h/2)\mu(h/2)+\bm{q}^-\cdot \bm{x}(\bm{R},-h/2)\mu(-h/2)\\
=&h\bar{\bm{q}}\cdot\bm{r}+h\bm{m}\cdot\bm{y}_1+\frac{1}{12}h^3\bar{\bm{q}}\cdot\bm{y}_2+\frac{1}{24}h^3\bm{m}\cdot\bm{y}_3+O(h^4),
\end{split}
\end{align}
where we have  introduced the notations
\begin{align}\label{eq:qm}
\bar{\bm{q}}=\frac{\mu(h/2)\bm{q}^++\mu(-h/2)\bm{q}^-}{h},\quad \bm{m}=\frac{\mu(h/2)\bm{q}^+-\mu(-h/2)\bm{q}^-}{2}.
\end{align}
In a similar spirit, the load potential per unit arclength due to the lateral traction has the expansion
\begin{align}\label{eq:ql}
\int_{-h/2}^{h/2} \bm{q} \cdot \bm{x}(S,Z)\mu^*(Z)\,dZ=\bm{p}_0\cdot\bm{r}+\bm{p}_1\cdot\bm{y}_1+O(h^4),
\end{align}
where  the coefficients $\bm{p}_i, (i=0,1)$ are given by
\begin{align}
&\bm{p}_0=h(1+\frac{1}{24}h^2c_\tau^2)\bm{q}^{(0)}+\frac{1}{24}h^3(\bm{q}^{(2)}- 2\kappa_\tau \bm{q}^{(1)}),\quad \bm{p}_1=\frac{1}{12}h^3(\bm{q}^{(1)}-\kappa_\tau \bm{q}^{(0)}),
\end{align}
and $\bm{q}^{(i)}, (i=0,1,2)$ are  coefficients appearing in the expansion  $\bm{q}=\bm{q}^{(0)}+Z\bm{q}^{(1)}+\frac{1}{2}Z^2\bm{q}^{(2)}+O(Z^3)$.

Inserting the expansions \eqref{eq:EE}, \eqref{eq:qtb} and \eqref{eq:ql} into 3D energy functional \eqref{eq:E}, we obtain an approximate energy functional of the form
\begin{align}\label{eq:inter}
\begin{split}
&\mathcal{E}[\bm{r},\bm{y}_1,\bm{y}_2,\bm{y}_3]=\int_{\Omega} L\,dA-\int_{\partial\Omega_t}Q\,dS+O(A h^4),
\end{split}
\end{align}
where $A$ is the area of $\Omega$, and the quantities $L$ and $Q$ are given by
\begin{align}
\begin{split}\label{eq:G}
&L=h \bar{W}(\bm{F}_0)-h\bar{\bm{q}}\cdot\bm{r}-h\bm{m}\cdot\bm{y}_1+h^3 L_3,\\
&L_3=\frac{1}{12}  K \bar{W}(\bm{F}_0)+\frac{1}{24} \big(\bm{F}_1:\mathcal{A}:\bm{F}_1+2\bm{P}_0:( \nabla\bm{r} \bm{\kappa}^2+ \nabla \bm{y}_1\bm{\kappa})\\
&\quad\quad +\bm{P}_0\bm{n}\cdot \bm{y}_3-4H \bm{P}_0:\bm{F}_1\big)-\frac{1}{12}\bar{\bm{q}}\cdot\bm{y}_2-\frac{1}{24}\bm{m}\cdot\bm{y}_3,\\
&Q=\bm{p}_0\cdot\bm{r}+\bm{p}_1\cdot\bm{y}_1.
\end{split}
\end{align}

\subsection{Optimal director fields}

By the principle of stationary potential energy, one needs to find $\bm{r}, \bm{y}_1, \bm{y}_2$ and $\bm{y}_2$ such that the energy \eqref{eq:G} is stationary. This task can be made easier using a two-step variational procedure stemming from the variational asymptotic method \citep{berdichevskii1979variational}. First, we fix $\bm{r}$ and searching for $\bm{y}_1, \bm{y}_2$ such that the energy functional \eqref{eq:inter} remains stationary;  also, there is no need to consider the third-order director $\bm{y}_3$, since it 
does not actually enter the asymptotic expansion of the energy up to order $h^3$, which will be shown later. This calculation results in an optimal shell energy, which  depends only on $\bm{r}$.

When $\bm{r}$ is stipulated,  the optimal $\bm{y}_1$ and $\bm{y}_2$ satisfy the Euler--Lagrange equations
\begin{align}\label{eq:dG}
\frac{\partial L}{\partial\bm{y}_1}-\nabla\cdot\Big(\frac{\partial L}{\partial\nabla\bm{y}_1}\Big)^T=0,\quad \frac{\partial L}{\partial\bm{y}_2}=0,
\end{align}
where $\nabla\cdot (\bullet)=\bm{g}^\alpha\cdot (\bullet)_{,\alpha}$ denotes the surface divergence. Written out explicitly, these two equations take the form
\begin{align}
&\bar{W}_{\bm{F}}(\nabla\bm{r}+\bm{y}_1\otimes \bm{n})\bm{n}+h^2\Big[\frac{\partial L_3}{\partial\bm{y}_1}-\nabla\cdot\Big(\frac{\partial L_3}{\partial\nabla\bm{y}_1}\Big)^T\Big] =\bm{m}, \label{eq:mt}\\
&\bm{B}\bm{y}_2+ [\mathcal{A}:(\nabla\bm{r}\bm{\kappa}+\nabla\bm{y}_1)]\bm{n}-2H\bm{P}_0\bm{n}=\bar{\bm{q}},\label{eq:qt}
\end{align}
where $\bar{W}_{\bm{F}}=\frac{\partial\bar{W}}{\partial\bm{F}}$, and $\bm{B}$ is the acoustic tensor based on $\bm{n}$  defined by 
\begin{align}\label{eq:C}
\bm{B}\bm{a}:=(\mathcal{A}:\bm{a}\otimes\bm{n})\bm{n},\quad \forall \  \bm{a}\in\mathbb{R}^3,
\end{align}
which is equivalent to $B_{ij}=\mathcal{A}_{i3j3}$ in  component form.

Notice that when the terms of order $h^2$ is neglected,  Eq. \eqref{eq:mt} reduces to the simple form
\begin{align}\label{eq:WFF}
\bar{W}_{\bm{F}}(\nabla\bm{r}+\bm{y}_1\otimes \bm{n})\bm{n}=\bm{m}.
\end{align}
The strong ellipticity condition together with the implicit function theorem guarantees that  $\bm{y}_1$ can be uniquely solved in terms of $\nabla \bm{r}$ from \eqref{eq:WFF}. We denote this solution by $\bm{y}_1=\bm{f}(\nabla\bm{r})$ which satisfies
\begin{align}\label{eq:mm}
\bar{W}_{\bm{F}}(\nabla\bm{r}+\bm{f}(\nabla\bm{r})\otimes \bm{n})\bm{n}=\bm{m}.
\end{align}
and is an approximation solution to \eqref{eq:mt} correct to order $h$. A key observation (proved in \ref{app:x1})  is that , it is still asymptotically consistent to replace $\bm{y}_1$ by this membrane approximation $\bm{f}(\nabla\bm{r})$ at the present order of approximation. Thus in the subsequent derivation, we will use $\bm{y}_1=\bm{f}(\nabla\bm{r})$. Eq. \eqref{eq:qt} is linear in $\bm{y}_2$ and can be solved explicitly as:
\begin{align}\label{eq:g}
	\begin{split}
		\bm{y}_2=&\bm{B}^{-1}\big[ 2H\bm{P}_0\bm{n}-[\mathcal{A}:(\nabla\bm{r}\bm{\kappa}+\nabla\bm{y}_1)]\bm{n}+\bar{\bm{q}}\big]=\bm{g}(\nabla \bm{r},\nabla\nabla \bm{r}).
	\end{split}
\end{align}
where the last equation serves to define the function $\bm{g}$ and  $\nabla\nabla\bm{r}:=(\nabla\bm{r})_{,\alpha}\otimes\bm{g}^\alpha$ means the surface gradient of $\nabla\bm{r}$. It is noted that the dependence of $\bm{y}_2$ on $\nabla\bm{r}$ arises from the curvature tensor, and in particular, when the curvature vanishes (i.e., a plate), the optimal $\bm{y}_2$ is  a function of $\nabla\nabla\bm{r}$ only.

\subsection{The optimal shell energy}

Substituting the expressions of the optimal $\bm{y}_1$ and $\bm{y}_2$ (\eqref{eq:WFF} and \eqref{eq:g}) back into \eqref{eq:G}, after some simplification, we have
\begin{align}\label{eq:G1}
\begin{split}
L=&h(1+\frac{1}{12}h^2K) \bar{W}(\bm{F}_0)+\frac{1}{24}h^3(\bm{\rho}:\mathcal{A}:\bm{\rho}-\bm{y}_2\cdot\bm{B}\bm{y}_2-2\bm{P}_0\bm{1}:\bm{\rho}\bm{\kappa}^*),\\
&-h\bar{\bm{q}}\cdot\bm{r}-h\bm{m}\cdot\bm{y}_1+\underbrace{\frac{1}{24}h^3(\bm{P}_0\bm{n}-\bm{m})\cdot \bm{y}_3}_{=0},
\end{split}
\end{align}
where  $\bm{\rho}=(\nabla\bm{r})\bm{\kappa}+\nabla\bm{y}_1$
 is called the {\it bending strain tensor} describing the change in the curvature of the middle surface (\cite{steigmann2023lecture}, p. 126). The last term vanishes since $\bm{P}_0=\bar{W}_{\bm{F}}(\bm{F}_0)$ and $\bar{W}_{\bm{F}}(\bm{F}_0)-\bm{m}=0$  (see \eqref{eq:mm}), which justifies our earlier assertion that  $\bm{y}_3$ does not appear in the expansion of the energy up to order $h^3$. Applying (\ref{eq:G1}),  the 2D energy functional \eqref{eq:inter} simplifies to
\begin{align}\label{eq:twod}
\begin{split}
\mathcal{E}_\text{2d}[\bm{r}]=&\int_{\Omega} 
\Big(h(1+\frac{1}{12}h^2K)\bar{W}(\nabla\bm{r}+\bm{y}_1\otimes\bm{n})+\frac{1}{24}h^3(\bm{\rho}:\mathcal{A}:\bm{\rho}-\bm{y}_2\cdot \bm{B}\bm{y}_2-2 \bm{P}_0\bm{1}:\bm{\rho}\bm{\kappa}^*)\\
&-h\bar{\bm{q}}\cdot\bm{r}-h\bm{m}\cdot\bm{y}_1\Big)\,dA-\int_{\partial\Omega_t} (\bm{p}_0\cdot\bm{r}+\bm{p}_1\cdot \bm{y}_1)\,dS.
\end{split}
\end{align}
One can check that \eqref{eq:twod} reduces to Eq. (5.92) in \cite{steigmann2023lecture} when the material is linear,  the growth tensor is identity and the tractions on the top and bottom surfaces vanish.

The morphoelastic shell energy  \eqref{eq:twod} has the following nice interpretation: the first term $h(1+\frac{1}{12}h^2K)\bar{W}(\nabla\bm{r}+\bm{y}_1\otimes\bm{n})$ corresponds to the energy of  stretching strains, the second term $\frac{1}{24}h^3\bm{\rho}:\mathcal{A}:\bm{\rho}$ accounts for the contribution of the bending strains, and the fourth term $-\frac{1}{12}h^3\bm{P}_0\bm{1}: \bm{\rho}\bm{\kappa}^*$ arises from the coupling of the stretching, bending and curvature; the remaining terms are the load potential. The effect of growth is implicitly accounted for in the augmented energy function $\bar{W}$. We observe the third term $-\frac{1}{24}h^3\bm{y}_2\cdot \bm{B}\bm{y}_2$, correction to the bending energy to restore self-consistency, is frequently neglected in the existing literature. Our model encapsulates this term through an asymptotically consistent derivation.

The morphoelastic shell energy, as formulated in our work, maintains asymptotic consistency and adeptly accommodates finite stretching and bending, as well as their coupling effects. This energy functional is designed for straightforward integration into numerical simulations (see Section \ref{sec:example} for an example), offering improved computational efficiency in comparison to traditional three-dimensional computational frameworks.

\subsection{Equilibrium equations and boundary conditions}
We have obtained optimal two-dimensional shell energy by fixing the leading order coefficient $\bm{r}$ and optimizing the higher order coefficients $\bm{y}_1$ and $\bm{y}_2$. Next, we further optimize the energy functional with respect to $\bm{r}$ to extract the shell equations and relevant boundary conditions (detailed in \ref{app:B}). This incremental optimization offers an advantage over simultaneous optimization in  derivation of shell equations and boundary conditions since it is easier to use the results of the first step to do the simplification.
 The results for the Euler--Lagrange equation of equilibrium and the relevant boundary conditions are 
\begin{align}
&\nabla \cdot \bm{N}^T+h\bar{\bm{q}}=0\quad \text{in}\ \Omega, \label{eq:pde}\\
& \bm{N}\bm{\nu}-(\mathcal{M}[\bm{\tau},\bm{\nu}])_{,S}=\bm{p}_0-(\mathcal{F}^{T_{231}}[\bm{\tau},\bm{p}_1])_{,S}\quad \text{on}\ \partial\Omega_t, \label{eq:pde1}\\
&\mathcal{M}[\bm{\nu},\bm{\nu}]=\mathcal{F}^{T_{231}}[\bm{\nu},\bm{p}_1]\quad \text{on}\ \partial\Omega_t,\label{eq:M}\\
&\bm{r}=\bm{d},\quad \bm{r}_{,\nu}=\bm{\varphi}\quad  \text{on}\ \partial\Omega_u. \label{eq:u}
\end{align}
In the above expressions, the operations $\mathcal{C}[\bm{a},\bm{b}]=(\mathcal{C}\cdot \bm{b})\cdot\bm{a}$, $\bm{r}_{,\nu}=(\nabla\bm{r})\bm{\nu}$,  $\bm{d}$ are $\bm{\varphi}$ are prescribed vector-valued functions on the displacement edge, $\bm{N}$ is a second-order tensor and $\mathcal{M}$ and $\mathcal{F}$ are third-order tensors defined by
\begin{align}\label{eq:NMN}
&\bm{N}=\frac{\partial L}{\partial\nabla\bm{r}}-\nabla\cdot \mathcal{M}^{T_{312}},\quad \mathcal{M}=\frac{\partial L}{\partial\nabla\nabla\bm{r}},\quad \mathcal{F}=\frac{\partial\bm{f}}{\partial\nabla\bm{r}},
\end{align}
where $L$ is given in \eqref{eq:G1} with the last term dropped and the superscripts like ${T_{312}}$ denote the transpose of a third-order tensor \citep{pan2014tensor}. In particular, we have $(\mathcal{M}^{T_{312}})_{ijk}=\mathcal{M}_{kij}$ and $(\mathcal{F}^{T_{231}})_{ijk}=\mathcal{F}_{jki}$. Note that \eqref{eq:u} is valid for a clamped edge, and for a pinned edge, it should be modified: $\bm{r}$ is assigned on $\partial\Omega_u$ and \eqref{eq:M} holds on $\partial\Omega_u$  \citep{steigmann2023lecture}.

To summarize, the fourth-order nonlinear partial differential equation \eqref{eq:pde} together with the boundary conditions \eqref{eq:pde1}--\eqref{eq:u} constitute the  differential governing system for $\bm{r}$  (the  position of the actual middle surface). Once $\bm{r}$ is determined, the three-dimensional deformation is recovered by using the recurrence relations \eqref{eq:WFF} and \eqref{eq:g}.

We close this section with the following important remark. The above two-step variational procedure is rigorous and does not sacrifice any accuracy. It is mathematically equivalent to applying variations with respect to $\bm{r}$, $\bm{y}_1$ and $\bm{y}_2$ simultaneously. This equivalence may be illustrated by the following simple example.
Suppose that $f(x,y)$ is a function that we need to minimize. First we assume that $x$ is fixed and seek $y$ such that $f(x,y)$ is minimal, which implies the optimal $y=y_\text{opt}(x)$ satisfies $f_y(x,y_\text{opt}(x))=0$, where the subscript means partial differentiation. Inserting this into $f(x,y)$ and minimizing the resulting function $g(x):=f(x,y_\text{opt}(x))$, we obtain
\begin{align}
g'(x)=f_x(x,y_\text{opt}(x))+f_y(x,y_\text{opt}(x))y'_\text{opt}(x)=f_x(x,y_\text{opt}(x))=0,
\end{align} 
The solution $x_0$ to the above equation satisfies $f_x(x_0,y_\text{opt}(x_0))=f_y(x_0,y_\text{opt}(x_0))=0$, justifying the equivalence between the two-step minimization procedure and the approach of  simultaneous minimization. Readers are referred to  \cite{audoly2021asymptotic} for rigorous justification of the claim for the variational case.

\section{Connection with other reduced models}\label{sec:connection}
In this section, we show that our morphoelastic shell model  recovers other reduced models in literature. By applying the correct scaling assumptions, we can derive the classical Kirchhoff and Föppl-von Kármán shell theories from our more comprehensive model.
Our shell model is also  consistent with the finite-strain shell model of \cite{song2016consistent} that is derived from the three-dimensional differential equations. Besides, in the plane strain case for the localized necking, our shell model reduces to the one-dimensional gradient model of \cite{audoly2016analysis}. Since these models treat traditional elastomers, we take $\bm{G}=\bm{I}$ in this section.

\subsection{Small strain: justification of linear shell theory}

Let us consider that the shell has a linear isotropic constitutive law and the strains are small. For the small-strain problems, it is convenient to use a formulation in terms of displacement  $\bm{u}=\bm{x}-\bm{X}$. In the linear isotropic case, the strain energy function has the expression 
\begin{align}\label{eq:Ww}
W(\bm{F})=\frac{1}{2} \lambda (\tr (\bm{H}))^2+\mu \tr (\sym(\bm{H})^2),
\end{align}
where $\bm{H}=\frac{\partial\bm{u}}{\partial\bm{X}}=\bm{F}-\bm{I}$ is the gradient of the displacement and $\sym(\bm{H})=\frac{1}{2}(\bm{H}+\bm{H}^T)$ signifies the symmetric part of $\bm{H}$, and $\lambda$ and $\mu$ are the Lam\'{e} parameters. 

From the expansions of $\bm{x}$, we see that $\bm{u}$ can be expanded as 
\begin{equation}
\begin{aligned}
\bm{u}&=\bm{r}+ \bm{y}_1 Z+\frac{1}{2} \bm{y}_2\left(Z^2-\frac{h^2}{12}\right)+\cdots-(\bm{R}+Z\bm{n})\\
&=\bm{u}_0+\bm{u}_1 Z+\frac{1}{2}\bm{u}_2\Big(Z^2-\frac{1}{12}h^2\Big)+\cdots,
\end{aligned}
\end{equation}
where $\bm{u}_0=\bm{r}-\bm{R}$, $\bm{u}_1=\bm{y}_1-\bm{n}$ and $\bm{u}_2=\bm{y}_2$. We adopt the common assumption in linear shell theories that the traction $\bm{q}^+$ and $\bm{q}^-$ are of order $h^3$ \citep{steigmann2023lecture}. This implies that $\bar{\bm{q}}=O(h^2)$ and $\bm{m}=O(h^3)$. With higher-order terms neglected, Eqs. \eqref{eq:WFF} and \eqref{eq:g} yield the following expressions for the optimal directors
\begin{align}\label{eq:u1u2}
\bm{u}_1=-\bm{n}\cdot\nabla\bm{u}_0-\eta \tr(\bm{1}\nabla \bm{u}_0)\bm{n},\quad \bm{u}_2=-\bm{n}\cdot\nabla\bm{\rho}-\eta \tr(\bm{1}\nabla \bm{\rho})\bm{n},
\end{align}
where $\eta=\frac{\lambda}{\lambda+2\mu}$ and $\bm{\rho}=(\nabla\bm{u}_0)\bm{\kappa}+\nabla\bm{u}_1$ is the bending strain tensor which can be further simplified into the form (see \ref{app:bending})
\begin{equation}
\label{Eq_bending_strain}
\bm{\rho}=-\bm{1}(\bm{n}\cdot\nabla\nabla\bm{u}_0)+(2\bm{\varepsilon}+\eta \tr(\bm{\varepsilon})\bm{1})\bm{\kappa}-\eta\bm{n}\bm{\otimes}\nabla\tr(\bm{\varepsilon}).
\end{equation}

In view of linear constitutive law \eqref{eq:Ww} and the expressions of the optimal directors \eqref{eq:u1u2}, we find that the two-dimensional shell energy functional \eqref{eq:twod} takes the form
\begin{align}
\begin{split}\label{eq:shell}
\mathcal{E}_{\text{shell}}[\bm{u}_0]=&\int_{\Omega}\Big(h (1+\frac{1}{12}h^2K)[\eta \mu (\tr(\bm{\varepsilon}))^2+\mu \tr(\bm{\varepsilon}^2)]+\frac{1}{12}h^3[\eta \mu (\tr(\bm{\rho}_t))^2\\
&+\mu \tr(\sym(\bm{\rho}_t)^2)-2\eta\mu\tr(\bm{\varepsilon})\tr(\bm{\rho}_t\bm{\kappa}^*)-2\mu\tr(\bm{\varepsilon}\bm{\rho}_t\bm{\kappa}^*)]-h\bar{\bm{q}}\cdot\bm{u}_0\Big)\,dA,
\end{split}
\end{align}
where we have ignored the boundary terms and $\bm{\rho}_t=\bm{1}\bm{\rho}$ denotes the in-plane part of $\bm{\rho}$ given by
$
\bm{\rho}_t=-\bm{1}(\bm{n}\cdot\nabla\nabla\bm{u}_0)+(2\bm{\varepsilon}+\eta \tr(\bm{\varepsilon})\bm{1})\bm{\kappa}
$
and  $\bm{\varepsilon}=\frac{1}{2}(\bm{1}\nabla\bm{u}_0+(\bm{1}\nabla\bm{u}_0)^T)$. The above form of shell energy (\ref{eq:shell}) is consistent with that of the linear Kirchhoff shell theory given in textbooks \citep{ventsel2002thin,reddy2006theory}. A similar argument shows that the F\"{o}ppl-von K\'{a}rm\'{a}n type of shell theory can also be recovered from our shell model: one starts with constitutive law \eqref{eq:Ww}  with $\bm{H}$ being replaced by the Lagrange--Green strain tensor $\bm{E}=\frac{1}{2}(\bm{H}+\bm{H}^T)+\bm{H}^T\bm{H}$ and only retains nonlinear terms of the form $u_{03,\alpha}u_{03,\beta}$ with $u_{03}=\bm{u}_0\cdot\bm{n}$.

In the particular case where the shell is a plate, setting $\bm{\kappa}=0$ in \eqref{eq:shell}, we recover the energy functional of the classical Kirchhoff  plate theory
\begin{align}
\begin{split}
\mathcal{E}_\text{plate}[\bm{u}_0]=& \int_{\Omega} \Big( h[\eta \mu (\tr(\nabla \bm{u}_{0t}))^2+\mu \tr(\sym(\nabla \bm{u}_{0t})^2) ]\\
&+\frac{1}{12}h^3  [\eta \mu (\tr(\nabla\nabla u_{03}))^2+\mu \tr((\nabla\nabla u_{03})^2)] -h\bar{\bm{q}}\cdot\bm{u}_0\Big)\,dA,
\end{split}
\end{align}
where we used the decomposition $\bm{u}_0=\bm{u}_{0t}+u_{03} \bm{n}$ with $\bm{u}_{0t}=\mathbf{1} \bm{u}$.

In the special case of pure bending $\bm{\varepsilon}=0$, we have simplified results. In this case, the bending strain tensor becomes $\bm{\rho}=-\bm{1}(\bm{n}\cdot\nabla\nabla\bm{u}_0)$. It is well-known in the surface elasticity theory that the term $\bm{1}(\bm{n}\cdot\nabla\nabla\bm{u}_0)$ is exactly equal to the variation of the curvature of the middle surface \citep{steigmann1999elastic}. Thus we have
\begin{align}\label{Eq4_6}
-\rho_{\alpha\beta}=b_{\alpha\beta}-\kappa_{\alpha\beta},
\end{align}
where $b_{\alpha\beta}$ and $\kappa_{\alpha\beta}$ are components of the curvatures tensors of the actual and initial middle surfaces with respect to the contravairant basis $\{\bm{g}^1,\bm{g}^2\}$, respectively. Eq. \eqref{Eq4_6} clearly shows the geometric meaning of $\bm{\rho}$.

\subsection{Finite strain: equivalence with differential equation-based shell theory}
In this part, we want to show that our shell model is consistent with that of Dai {\it et al.}'s \citep{song2016consistent} which were derived from 3D differential equations. We first review Dai's shell theory, and then do the comparison.

In a series of papers \citep{dai2014consistent,song2016consistent,yu2020refined}, Dai and coworkers derived a consistent plate/shell theory for finite-strain deformations, starting from the three-dimensional governing equations. Their derivation is built on series expansions
\begin{equation}
\begin{aligned}\label{eq:xx}
\bm{x}&=\bm{x}^{(0)}+\bm{x}^{(1)}Z+\frac{1}{2}\bm{x}^{(2)}Z^2+\frac{1}{6}\bm{x}^{(3)}Z^3+\cdots,\\
\bm{F}&=\bm{F}^{(0)}+\bm{F}^{(1)}Z+\frac{1}{2}\bm{F}^{(2)}Z^2+\cdots,\\
\bm{P}&=\bm{P}^{(0)}+\bm{P}^{(1)}Z+\frac{1}{2}\bm{P}^{(2)}Z^2+\cdots.
\end{aligned}
\end{equation}
The equilibrium equation of the shell theory can be written as
\begin{align}\label{eq:diff}
\nabla\cdot\bar{\bm{P}}^T+\bar{\bm{q}}=0,
\end{align}
where $\bar{\bm{P}}$ is the stress resultant defined by
\begin{align}
\bar{\bm{P}}=\frac{1}{h}\int_{-\frac{h}{2}}^{\frac{h}{2}} \bm{P}(\bm{1}-Z\bm{\kappa}^*)\,dZ+O(h^4)=\bm{P}^{(0)}\bm{1}-\frac{1}{12}h^2 \bm{P}^{(1)}\bm{\kappa}^*+\frac{1}{24}h^2\bm{P}^{(2)}\bm{1}.
\end{align}
According to the expression of $\bar{\bm{P}}$, three recurrence relations for $\bm{x}^{(1)}, \bm{x}^{(2)}$ and $\bm{x}^{(3)}$ are required, which can be found by substituting the expansion \eqref{eq:xx} into 3D equilibrium equations and solving the two resulting equations by evaluating the coefficients of $Z^0$ and $Z^1$ as well as one equation obtained from the bottom and top traction conditions. Then the equilibrium equation \eqref{eq:diff} supplemented by the recurrence relations  contains  the leading-order expansion coefficient $\bm{x}^{(0)}$ only, which is the shell equation. The 2D boundary conditions are derived through averaging  the three-dimensional ones separately.

We now show that our shell equilibrium equation is equivalent to that of \cite{song2016consistent}.  
Starting from \eqref{eq:xx} and repeating the dimension reduction procedure in Section \ref{sec:shell}, we  eventually arrive at an approximate energy functional of the form 
\begin{align}\label{eq:2dd}
\begin{split}
\mathcal{E}[\bm{x}^{(0)},\bm{x}^{(1)},\bm{x}^{(2)},\bm{x}^{(3)}]=&\int_{\Omega} \Big(h(1+\frac{1}{12}h^2K)\bar{W}(\bm{F}^{(0)})+\frac{1}{24}h^3(\bm{P}^{(0)}:\bm{F}^{(2)}+\bm{P}^{(1)}:\bm{F}^{(1)}\\\
&-4H\bm{P}^{(0)}:\bm{F}^{(1)})-h\bar{\bm{q}}\cdot\bm{x}^{(0)}-h\bm{m}\cdot\bm{x}^{(1)}-\frac{1}{8}h^3\bar{\bm{q}}\cdot\bm{x}^{(2)}\\
&-\frac{1}{24}h^3\bm{m}\cdot\bm{x}^{(3)}\Big)\,dA+\text{boundary terms}.
\end{split}
\end{align} 
The variations with respect to $\bm{x}^{(1)}$, $\bm{x}^{(2)}$, and $\bm{x}^{(3)}$ lead to the recurrence relations that are asymptotically consistent with those of \cite{song2016consistent}. Taking the recurrence relations into the above functional energy  \eqref{eq:2dd},  
it is direct to check that the variation of it with respect to $\bm{x}^{(0)}$ yields the equilibrium equation \eqref{eq:diff}, which proves the consistency of the current shell model with that of \cite{song2016consistent} derived from 3D differential equations.

In comparing our shell model to that given by \cite{song2016consistent}, we recognize that except the consistency of the two models, they also have their own distinctive features. 
 Our shell model is user-friendly for numerical simulations, providing an optimal 2D shell energy that is based on the position of the actual middle surface.
In contrast, the shell theory developed by Dai {\it et al.} is articulated through 2D shell equations that are formulated using stress components and based on the position of the deformed middle surface, which offers more direct physical interpretations, making it particularly advantageous for equation-driven analysis or analytical investigations.
It is important to highlight that the average boundary conditions employed by Dai {\it et al.}'s, while being reasonable two-dimensional approximations, may sacrifice a degree of consistency with the original variational principles. Our boundary conditions differ and aim to maintain closer alignment with the original variational structure.

\subsection{Plane strain case: equivalence with one-dimensional gradient model}

\begin{figure}[h!]
	\centering
	\includegraphics[width=0.9\linewidth]{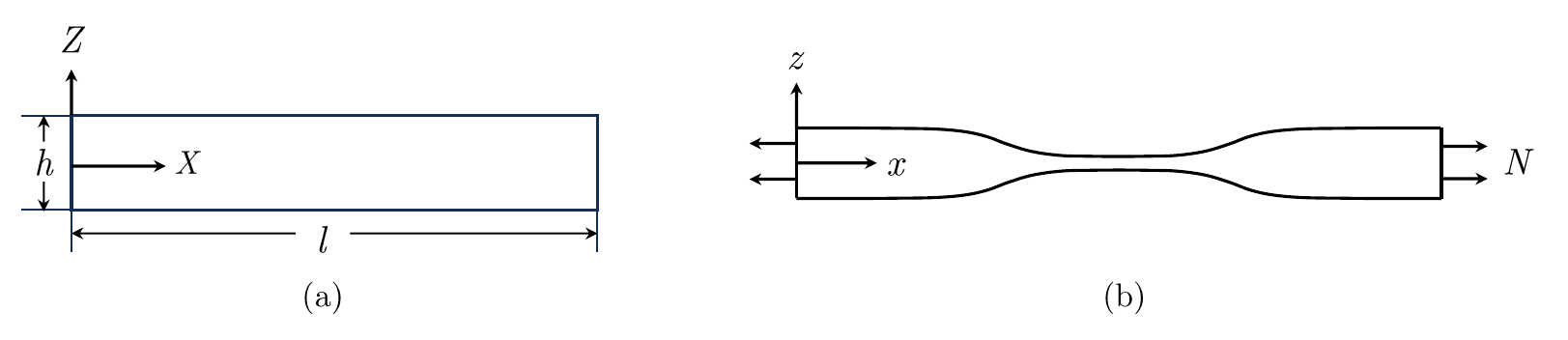}
	\caption{Plane strain case: (a) a plate with length $l$ and thickness $h$ in the reference configuration; (b) the plate 
		is subjected to the axial force at two ends, and undergoes localized necking.  }
	\label{fig:block}
\end{figure}

We consider a plate (flat shell) with initial length $l$ and thickness $h$, see Fig. \ref{fig:block}(a). Suppose it is subjected to the 
axial force $N$ (per unit cross-sectional area)  at its two ends, and the top and bottom surfaces remain traction-free. It then
 undergoes a localized necking, which is symmetric about $X$-axis, and is regarded as a plane-strain deformation, see Fig. \ref{fig:block}(b).

The material point with position $(X,Z)$ in the reference configuration is transported to the position $(x(X,Z),z(X,Z))$ in the current configuration, where the Cartesian basis is used.  In view of the symmetry of the deformation, one can expand the even function $x(X,Z)$ and odd function $z(X,Z)$  as
\begin{align}
&x(X,Z)=x_0(X)+\frac{1}{2}x_2(X)\Big(Z^2-\frac{1}{12}h^2\Big)+O(h^4),\\
&z(X,Z)=z_1(X)Z+O(h^3).
\end{align}

Let's posit that the constitutive behavior of the shell is characterized by a strain energy function, $W(\lambda_1,\lambda_2)$, where $\lambda_1$ and $\lambda_2$ represent the two principal stretches  align with the $X$ and $Z$ directions for homogeneous deformations. Then  \eqref{eq:WFF} and \eqref{eq:g} imply that
\begin{align}\label{eq:W2x2}
z_1=z_1(x'_0), \quad x_2=-\frac{z_1 z_1'}{x'_0},
\end{align}
where $z_1$ is an implicit function of $x'_0$ satisfying $W_2(x_0',z_1)=0$ with $W_2=\partial W/\partial \lambda_2$, and the prime means the derivative with respect to $X$. With the substitution of these relations,  the shell energy \eqref{eq:twod} can be written as
\begin{align}
\mathcal{E}_\text{1d}[x_0]=\int_0^l\Big(W(x'_0,z_1(x'_0))+\frac{h^3}{24}W_1(x'_0,z_1(x'_0))\frac{p(x'_0)^2}{x'_0}x_0''(X)^2-Nx'_0\Big)\,dX,
\end{align}
where $W_1=\partial W/\partial\lambda_1$ and $p(x_0')=dz_1/dx_0'$. One can verify that this energy agrees with that of one-dimensional model of \cite{audoly2016analysis}  for a two-dimensional block  by correlating $(\nu_0,\mu_0,\lambda,n_0,n)$  from their Eq. (2.29c) with our parameters $(-\frac{x'_0p}{z_1},z_1,x'_0,W_1,N)$, respectively.

\section{Comparison with  three-dimensional benchmark problems}\label{sec:application}
In this section, we validate our morphoelastic shell model by two 3D benchmark problems with exact solutions. 
 The axisymmetric deformation of a cylindrical shell and the large bending deformation of a flat shell  due to growth  are studied. These two benchmark problems admit exact solutions, which will be compared with the approximate solutions obtained from the present morphoelastic shell model.

\subsection{A growing cylindrical shell subjected to external traction}

We consider a thin-walled circular cylindrical shell, which in its reference configuration has inner radius $A=R_m+h/2$ and outer radius   $B=R_m-h/2$, where $R_m$ and $h$ are the radius of the middle surface and thickness of the shell, respectively.  Cylindrical coordinates $(R,\varTheta,X)$ and $(r,\theta,z)$ are employed in the reference and current configurations, respectively. The common set of standard basis vectors associated with both the reference and current cylindrical coordinates is denoted by $\{\bm{e}_r,\bm{e}_\theta,\bm{e}_z\}$.

Under the combined loading of a growth process and applied traction, denoted as $\bm{q}^+=q^+\bm{e}_r$ on its outer surface, the cylindrical shell experiences an axisymmetric  deformation.
 This deformation is described by
 \begin{align}\label{eq:rR}
 r=r(R),\quad \theta=\varTheta,\quad z=\lambda X,
 \end{align}
 where $\lambda$ is the axial stretch,  $A\leq R\leq B$ and $0\leq \varTheta\leq 2\pi$. Without loss of generality,  we set $\lambda=1$. The growth tensor is specified by
 \begin{align}
\bm{G}=\gamma\bm{e}_r\otimes\bm{e}_r+\gamma\bm{e}_\theta\otimes\bm{e}_\theta+\bm{e}_z\otimes\bm{e}_z,
 \end{align}
 where $\gamma$ is a constant.   It follows  that three principal elastic stretches in $r$-, $\theta$-, and $z$-directions are respectively
\begin{align}
\alpha_1=\frac{r'(R)}{\gamma},\quad \alpha_2=\frac{r}{R\gamma},\quad \alpha_3=1.
\end{align}

As in \cite{ogden1984non}, we adopt the following strain energy function for the elastic response of the biological shell:
\begin{align}\label{eq:WO}
W(\alpha_1,\alpha_2,\alpha_3)= \frac{2}{27}\nu (\alpha_1+\alpha_2+\alpha_3)^3-\nu(\alpha_1\alpha_2+\alpha_1\alpha_3+\alpha_2\alpha_3)+\nu,
\end{align}
where $\nu> 0$ is a material constant. Also, for notational simplicity, we scale all stress variables by $\nu$ and length variables by $R_m$, which is equivalent to setting $\nu=1$ and $R_m=1$, and we use the same letters to denote the scaled quantities. In particular, we  now have $A=1-h/2$ and $B=1+h/2$.
The non-zero components of the first P--K stress are given by
\begin{align}
P_{11}=\gamma \frac{\partial W}{\partial\alpha_1}, \quad P_{22}=\gamma\frac{\partial W}{\partial\alpha_2},\quad P_{33}=\gamma^2\frac{\partial W}{\partial\alpha_3}.
\end{align}
The equilibrium equation and the boundary conditions are
\begin{equation}
\label{Eq5_6}
\begin{aligned}
&\frac{dP_{11}}{dR}+\frac{P_{11}-P_{22}}{R}=0,\\
&P_{11}|_{R=A}=0,\quad P_{11}|_{R=B}=q^+.
\end{aligned}
\end{equation}
From (\ref{Eq5_6}), we obtain 
\begin{align}\label{eq:rr}
r(R)=C_1 R+\frac{C_2}{R}.
\end{align}
where
\begin{align}
&C_1=\frac{\gamma}{16}+\frac{3}{16}\sqrt{25\gamma^2+\frac{32B^2\gamma q^+ }{B^2-A^2}}, \quad C_2=\frac{A^2B^2q^+}{B^2-A^2}. \label{eq:C2}
\end{align}
Eqs. (\ref{eq:rr}) with (\ref{eq:C2}) are the exact solutions given by the three-dimensional morphoelasticty.

Then we turn to find the approximate solution given by the present shell model.  When the shell is circular cylindrical with the middle surface radius $R_m=1$, the coordinate along the thickness direction is identified with $Z=R-1$ so that $r(R)=r(1+Z)$ and the curvature tensor of the middle surface is given by $\bm{\kappa}=-\bm{e}_\theta\otimes\bm{e}_\theta$. This implies the mean and Gaussian curvatures are $H=-1/2$ and $K=0$, respectively. Accordingly, the determinant   $\mu(Z)=1-2HZ+KZ^2=1+Z$ and \eqref{eq:qm} implies that
\begin{align}
\bar{\bm{q}}=\frac{1+h/2}{h}q^+\bm{e}_r,\quad \bm{m}=\frac{1+h/2}{2}q^+\bm{e}_r.
\end{align}

Let $r_0=\frac{1}{h}\int_{-h/2}^{h/2} r(1+Z)\,dZ$ be the average of $r$ across the thickness. Expanding $r(1+Z)-r_0$ into a Taylor series at $Z=0$ yields
\begin{align}\label{eq:r}
r(1+Z)=r_0+r_1Z+\frac{1}{2}r_2 \Big(Z^2-\frac{1}{12}h^2\Big)+O(h^3).
\end{align}
From \eqref{eq:WFF} and \eqref{eq:g}, we find that the optimal $r_1$ and $r_2$ are given by
\begin{align}
&r_1=-r_0-\gamma+\frac{3}{2}\sqrt{\gamma(2r_0+2\gamma+qh)},\quad r_2=\frac{2r_0^2-4r_0r_1-6r_1^2+\gamma r_1+9\gamma q+7\gamma^2}{4(r_0+r_1+\gamma)}, \label{eq:r2}
\end{align}
where we have introduced the scale $q=q^+/h$. 

On substituting  \eqref{eq:r2} above into  \eqref{eq:twod} and solving the equation $d\mathcal{E}_\text{2d}/dr_0=0$, we obtain an approximate solution to $r_0$ correct up to order $h^2$:
\begin{align}
\begin{split}\label{eq:r0}
r_0=&\frac{1}{16}(8q+\gamma+3\sqrt{\gamma(16 q+25\gamma)})+\frac{3}{2}q\sqrt{\frac{\gamma}{16q+25\gamma}}h\\
&-\frac{5}{24}q\left(1-\frac{45\gamma^{3/2}}{(16q+25\gamma)^{3/2}}\right)h^2
+O(h^3).
\end{split}
\end{align} 
On the other hand,  it follows from \eqref{eq:rr} that the exact $r_0$ is given by
\begin{align}\label{eq:r0ex}
r_{0}^\text{ex}=\frac{1}{h}\int_{-h/2}^{h/2}r(1+Z)\,dZ=C_1+\frac{C_2}{h}\ln\frac{2+h}{2-h}.
\end{align}
With the use of  \eqref{eq:C2}, it is straightforward to check that expanding \eqref{eq:r0ex} with respect to $h$ to the order $h^2$ produces the same expression as \eqref{eq:r0}, validating that our shell model can deliver $O(h^2)$-correct results.

\subsection{Large bending of a flat shell induced by gradient growth}

\begin{figure}[htbp!]
	\centering
	\includegraphics[width=0.8\linewidth]{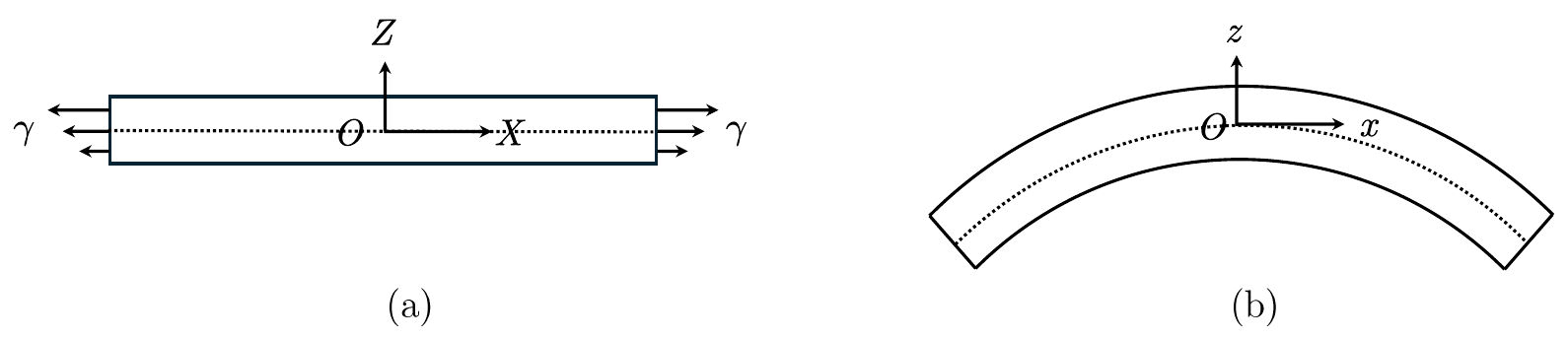}
	\caption{(a) Reference configuration of a plate; (b) illustration of the bending deformation of the plate induced by a gradient growth field. }
	\label{fig:bending}
\end{figure}

In this example, we analyze the large bending deformation of a plate (flat shell) induced by a gradient growth field, which is illustrated in Fig. \ref{fig:bending}. Here Cartesian coordinates $(X,Y,Z)$ and $(x,y,z)$ are used for a reference point and a current point, respectively.  The plate is of length $l$ (in the $X$-direction), thickness $h$ ($h \ll l$), and is subjected to  a pure growth field $\gamma(X, Y, Z)$. The growth field is supposed to be in the $X$-direction and varies linearly with $Z$:
\begin{align}\label{eq:gamm}
\gamma(X, Y, Z)=\gamma_0+\gamma_1 Z,
\end{align}
where $\gamma_0$ and $\gamma_1$ are two constants. To remove the rigid body motion, we set the middle point on the middle surface to be  fixed, and  the  deformation is symmetric with respect to $X=0$. The elastic response of the plate is assumed to be described by the compressible neo-Hookean model
\begin{align}\label{eq:neo-Hookean}
W=\frac{C}{2}(I_1-3-2\ln J)+\frac{D}{2}(J-1)^2,
\end{align}
where $C$ and $D$ are respectively the shear modulus and first Lame's constant, $I_1=\tr(\bm{A}^T\bm{A})$ and $J=\det(\bm{A})$. By scaling all stress variables by $C$, we may take $C=1$.

By solving the 3D governing system with a novel analytic approach, \cite{du2020analytical} showed that the deformation induced by the growth functions \eqref{eq:gamm} admits an exact solution which is of the plane-strain form with $y=Y$ and
\begin{align}\label{eq:xzex}
x(X,Z)=\Big(\frac{\gamma_0}{\gamma_1}+Z\Big)\sin(\gamma_1X),\quad z(X,Z)=\Big(\frac{\gamma_0}{\gamma_1}+Z\Big)\cos(\gamma_1X)-\frac{\gamma_0}{\gamma_1}.
\end{align}

To analyze this problem using the present shell model, we consider the expansions of $x(X,Z)$ and $z(X,Z)$ as follows: 
\begin{align}
\begin{split}
x(X,Z)&=x_0(X)+x_1(X)Z+\frac{1}{2}x_2(X)(Z^2-\frac{1}{12}h^2)+O(h^3),\\
z(X,Z)&=z_0(X)+z_1(X)Z+\frac{1}{2}z_2(X)(Z^2-\frac{1}{12}h^2)+O(h^3), 
\end{split}
\end{align}
where $x_0$ and $z_0$ are the averages of $x$ and $z$ across the thickness, respectively.
 For a plate structure, we have the curvature $\bm{\kappa}=0$. We also have $\bar{\bm{q}}=0$ and $\bm{m}=0$ 
 since upper and lower surfaces are traction-free. It then follows from \eqref{eq:WFF} that 
\begin{align}
\begin{split}
x_1&=-\frac{\gamma_0 z'_0(D\sqrt{x'^2_0+z'^2_0}+\sqrt{4\gamma_0^2+D(4+D)(x'^2_0+z'^2_0)})}{2\sqrt{x'^2_0+z'^2_0}(\gamma_0^2+D(x'^2_0+z'^2_0))},\\
z_1&=\frac{\gamma_0 x'_0(D\sqrt{x'^2_0+z'^2_0}+\sqrt{4\gamma_0^2+D(4+D)(x'^2_0+z'^2_0)}}{2\sqrt{x'^2_0+z'^2_0}(\gamma_0^2+D(x'^2_0+z'^2_0))}.
\end{split}
\end{align}
The expressions for $x_2$ and $z_2$ can be obtained similarly from \eqref{eq:g}, which are omitted for brevity. 

Substituting the above expressions into \eqref{eq:twod} and optimizing the resulting functional with respect to $x_0$ and $z_0$, we obtain a boundary-value problem of differential equations for $x_0$ and $z_0$.  At the leading-order, we have
\begin{align}
x'^2_0(X)+z'^2_0(X)=\gamma_0^2,
\end{align}
which implies that $x_0'$ and $z'_0$ are of the form
\begin{align}\label{eq:xz0p}
x'_0(X)=\gamma_0 \cos(a(X)),\quad z'_0(X)=-\gamma_0\sin(a(X)),
\end{align}
where $a(X)$ is a function to be determined. After substitution of the solution \eqref{eq:xz0p}, the next-order system becomes
 \begin{align}\label{eq:aa}
 \begin{split}
 &2(1+D)(2+D)^2\sin(a(X))a''(X)-(4+10D+15D^2+5D^3)\cos(a(X))a'(X)^2\\
 &+2D^2(7+3D)\gamma_1\cos(a(X))a'(X)+(4+10D+D^2-D^3)\gamma_1^2\cos(a(X))=0,\\
 &a(0)=0,\quad a'(l/2)=\gamma_1.
 \end{split}
 \end{align} 
It is observed that the solution to \eqref{eq:aa} is given by 
 \begin{align}\label{eq:a}
 a(X)=\gamma_1X.
 \end{align}
 In view of \eqref{eq:a}, integrating \eqref{eq:xz0p}  subjected to the boundary conditions $x_0(0)=0$ and $z_0(0)=0$ yields
 \begin{align}\label{eq:x0z0asy}
 x_0(X)=\frac{\gamma_0}{\gamma_1}\sin(\gamma_1 X),\quad z_0(X)=\frac{\gamma_0}{\gamma_1}(\cos(\gamma_1 X)-1).
 \end{align}
 
It is seen that this set of solutions (\ref{eq:x0z0asy}) is  exactly equal to the average of the exact solution \eqref{eq:xzex} across the thickness. This recovery of the exact solution provides a validation of the present shell model  for inhomogeneous deformation.

\section{Application to a biological example}\label{sec:example}

\begin{figure}[h!]
	\centering
	\subfloat[]{\includegraphics[width=0.32\textwidth]{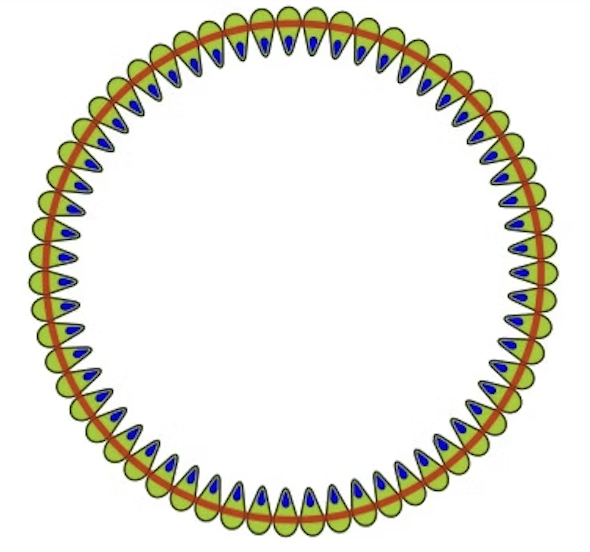}} \quad \quad \quad
	\subfloat[]{\includegraphics[width=0.32\textwidth]{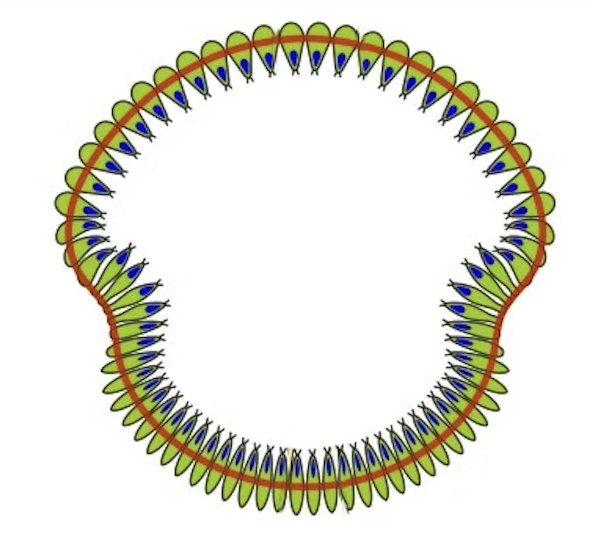} }
	\caption{The invagination of the cell sheet of the Volvox globator embryo. (a) The initial configuration and (b) the current configuration. Adapted from \cite{hohn2011there} under the terms of the Creative Commons Attribution License (http://creativecommons.org/licenses/by/2.0).} 
	\label{Fig5}
\end{figure}

In this section, we apply our morphoelastic shell model to explore a biological phenomenon: the invagination process in a monolayer cell sheet, exemplified by the embryo of Volvox globator. Initially, the cell sheet exhibits a spherical form (refer to Fig. \ref{Fig5}(a)). Subsequently, invagination occurs around the equatorial region (illustrated in Fig. \ref{Fig5}(b)), which is a precursor to the upward inversion of the southern hemisphere.
This process bears resemblance to critical stages in animal development, such as gastrulation, neurulation, and organogenesis \citep{keller2003we}, making it a subject worthy of detailed study. We acknowledge the valuable contributions of prior study by \cite{haas2021morphoelasticity}, who have developed a shell model  addressing large intrinsic curvatures within the context of small strains. Our current investigation,  extending to the finite strains, aims to provide extra interesting information about  the invagination mechanisms.

The initial configuration is a spherical shell with constant thickness $h$ and middle surface radius $R_m$. By scaling all length variables by $R_m$, we may take $R_m=1$. We set this biological shell experiences internally generated active stretches only and deforms axisymmetrically. The common set of cylindrical basis with both the reference and current cylindrical coordinates is denoted by $\{\bm{e}_r,\bm{e}_\theta,\bm{e}_z\}$. We also introduce 
a new set of curvilinear coordinates in the reference configuration, which are respectively the azimuthal angle, the arclength and the thickness variables, denoted as $(\varTheta, s, Z)$.
 
 A material point on the undeformed shell is represented as 
\begin{align}\label{eq:X6}
\bm{X}=\sin s \bm{e}_r+(1-\cos s)\bm{e}_z+Z \bm{n},\quad 0\leq s \leq \pi,\ -\frac{h}{2}\leq Z \leq  \frac{h}{2},
\end{align}
where $Z=0$ denotes the initial middle surface with $\bm{n}$ being its unit outward normal vector.
The associated covariant and contravariant basis of the middle surface are 
\begin{align}
\bm{g}_1=\sin s \,\bm{e}_{\theta}=\sin^2 s \, \bm{g}^1,\quad \bm{g}_2=\cos s \, \bm{e}_r+\sin s \,\bm{e}_z=\bm{g}^2,\quad \bm{g}_3=\bm{g}^3=\bm{n}.
\end{align}
The deformed shell is axisymmetric about $z$-axis, and we represent the current position of the material point by
\begin{align}\label{eq:X7}
\bm{x}=r(s,Z)\bm{e}_r+z(s, Z)\bm{e}_z.
\end{align}
The deformation gradient is then
\begin{align}
\bm{F}=\frac{\partial \bm{x}}{\partial \bm{X}}=\bm{A} \bm{G}.
\end{align}
We assume the growth of the cell sheet is mainly in the azimuthal and meridional directions and the growth tensor takes the form of \citep{li2023general} 
\begin{align}\label{eq:gg}
\begin{split}
&\bm{G}=\gamma_1(s,Z) \bm{g}_1\otimes\bm{g}^1+\gamma_2(s,Z) \bm{g}_2\otimes\bm{g}^2+\bm{n}\otimes\bm{n},\\
&\gamma_1(s,Z)=\gamma_1^{(0)}(s)+ Z \gamma_1^{(1)}(s),\quad 
\gamma_2(s,Z)=\gamma_2^{(0)}(s)+ Z \gamma_2^{(1)}(s),
\end{split}
\end{align}
where $\gamma_1^{(0)}, \gamma_1^{(1)}, \gamma_2^{(0)}$ and $\gamma_2^{(1)}$ are active stretches to be specified.

We assume that the elastic behavior of the biological shell is characterized by the compressible neo-Hookean material model given in \eqref{eq:neo-Hookean}  with $D=2C\nu/(1-2\nu)$, where $\nu\in[0,1/2]$ is Poisson's ratio. The incompressible model corresponds to $\nu\to 1/2$, which was considered in \cite{haas2021morphoelasticity}. The total potential energy of the morphoelastic shell is thus given by
\begin{align}\label{eq:3dd}
\mathcal{E}[r,z]= \int_0^{\pi}   \int_{-h/2}^{h/2}\det(\bm{G})W(\bm{F}\bm{G}^{-1})(1+Z)^2 2\pi\sin s \,d Z\,d s.
\end{align}
There is one dimension (the azimuthal angle variable) reduced  in (\ref{eq:3dd})  due to the axisymmetry of the deformation. 

By applying the present shell model, the current position $(r,z)$ are expanded into series as
\begin{align}
\begin{split}
r(s, Z)=r_0(s)+r_1(s) Z+\frac{1}{2}r_2(s)\Big(Z^2-\frac{1}{12}h^2\Big)+\cdots,\\
z(s, Z)=z_0(s)+z_1(s) Z +\frac{1}{2}z_2(s)\Big(Z^2-\frac{1}{12}h^2\Big)+\cdots.
\end{split}
\end{align}
Then based on the results of \eqref{eq:twod}, we obtain the optimal shell energy from \eqref{eq:3dd} as
\begin{align}\label{eq:1d}
\mathcal{E}_\text{1d}[r_0,z_0]=\int_0^{\pi} L(s,r_0,z_0,r'_0,z'_0,r_1,z_1,r'_1,z'_1)\,d s,
\end{align}
which is a one-dimensional energy functional. The $(r_1, z_1)$ involved  are  functions of $(r_0, z_0)$, satisfying the equation   \eqref{eq:mm}, which leads to
\begin{align}\label{eq:f1f2}
f_1(s,r_0,z_0,r'_0,z'_0,r_1,z_1)=0,\quad f_2(s,r_0,z_0,r_0',z'_0,r_1,z_1)=0.
\end{align}
The lengthy expressions of $L, f_1, f_2$ are not presented here. Due to the high nonlinearity, it seems not possible to solve  \eqref{eq:f1f2} for $r_1$ and $z_1$, and the Euler-Lagrange equation of (\ref{eq:1d}) analytically. In the following, we develop a numerical scheme to determine all the unknowns.

The Rayleigh–Ritz method \citep{dhatt2012finite} is employed to produce the numerical solution, which is a direct numerical methodfor minimizing a given functional. For easier implementation, $r_1$ and $z_1$ are regarded as independent variables with   \eqref{eq:f1f2} being the  constraints. By employing the method of Lagrange multipliers to eliminate these constraints, the task of minimizing the energy functional \eqref{eq:1d} is transformed into minimizing the subsequent Lagrange functional:\begin{align}
\label{Eq6_10}
\begin{split}
\tilde{\mathcal{E}}_\text{1d}[r_0,z_0,r_1,z_1,m_1,m_2]=&\int_{0}^{\pi} \Big[L(s,r_0,z_0,r'_0,z'_0,r_1,z_1, r'_1,z'_1)-m_1(s) f_1(s,r_0,z_0,r'_0,z'_0,r_1,z_1\\
&-m_2(s) f_2(s,r_0,z_0,r'_0,z'_0,r_1,z_1)\Big]\,d s,
\end{split}
\end{align}
where $m_1$ and $m_2$ are Lagrange multipliers enforcing the constraints in \eqref{eq:f1f2}. Then we divide the domain $[0,\pi]$ into $n$ equal intervals and approximate the functions $r_i, z_i, m_{i+1}$, ($i=0,1$) using linear interpolation in terms of their values at the node points. For instance, on the $j$-th interval,  $r_0(s)$ is approximated by
\begin{align}
r_0(s)=r_0^j+\frac{r_0^{j+1}-r_0^{j}}{\ell}(s-j\ell), \quad j\ell\leq s\leq (j+1)\ell,
\end{align} 
where $\ell=\pi/n$ is the mesh size and $r_0^j$ denotes the numerical approximation of $r_0(j\ell)$. A two-point Gaussian quadrature rule is used to calculate the integration (\ref{Eq6_10}) on each interval. Summing these integrations gives a discrete energy in terms of nodal values of $r_i, z_i, m_{i+1}$, ($i=0,1$) which are the unknowns to be solved. By setting the partial derivative of the discrete energy with respect to these nodal values to zero, and solving the resulting system of algebraic equations using the Newton-Raphson method with an appropriate initial guess, we obtain the numerical solution for the unknowns $(r_0, z_0)$ and $(r_1, z_1)$. This numerical scheme is implemented in the symbolic computation software \cite{wolfram2024mathematica} and it is found that taking $n=500$ yields sufficiently accurate results.
We have also conducted finite element simulations utilizing the UMAT subroutine within Abaqus. In the numerical calculations, we choose $\nu=0.4$ and $h=0.1$, and scale all stress variables by shear modulus $C$.

\begin{figure}[h!]
	\centering
	\subfloat[]{\includegraphics[width=0.36\textwidth]{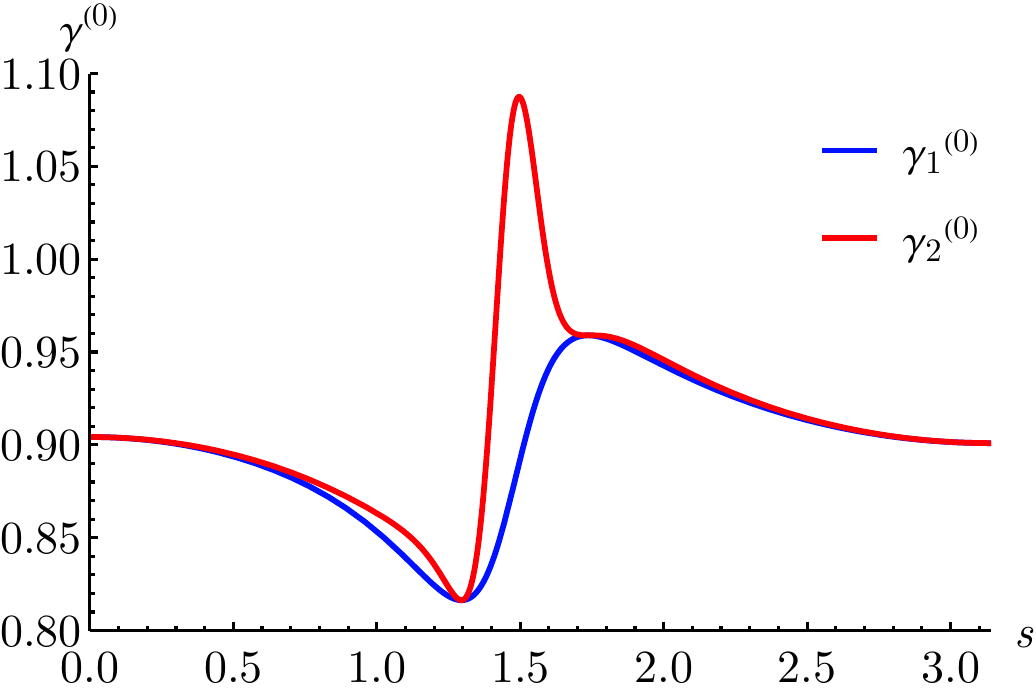}
	}\qquad\qquad\quad
	\subfloat[]{\includegraphics[width=0.36\textwidth]{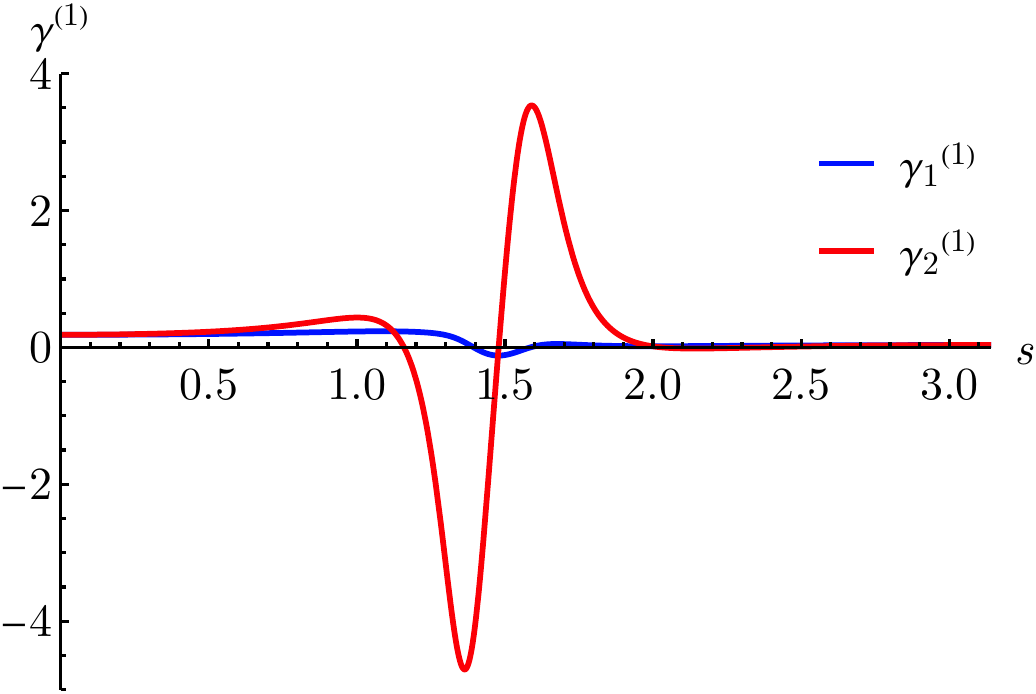}
	}
	\caption{(a) The leading-order active stretches. (b) The next-order  active stretches.}
	\label{fig:gamma}
\end{figure}

The growth functions defined in \eqref{eq:gg}  are determined using the approach given in \cite{li2023general}, where the authors proposed a systematic method to construct the growth functions based on the morphologies in the reference and current configurations. Due to lengthy expressions, the growth functions  are only shown  graphically in Fig. \ref{fig:gamma}. 
From the figure, one can see that, at the leading order, the cell sheet experiences circumferential active compression since $\gamma_1^{(0)}<1$. Examining the behavior of $\gamma_2^{(0)}$, one may find that  both the anterior and posterior regions of the cell sheet compress, whereas the equatorial region distinctly shifts from compression to expansion, and subsequently back to compression. This significant fluctuation of $\gamma_2^{(0)}$ in the equatorial zone is further emphasized by the modification of $\gamma_2^{(1)}$ at the equator. 

\begin{figure}[h!]
	\centering
	\includegraphics[width=0.9\textwidth]{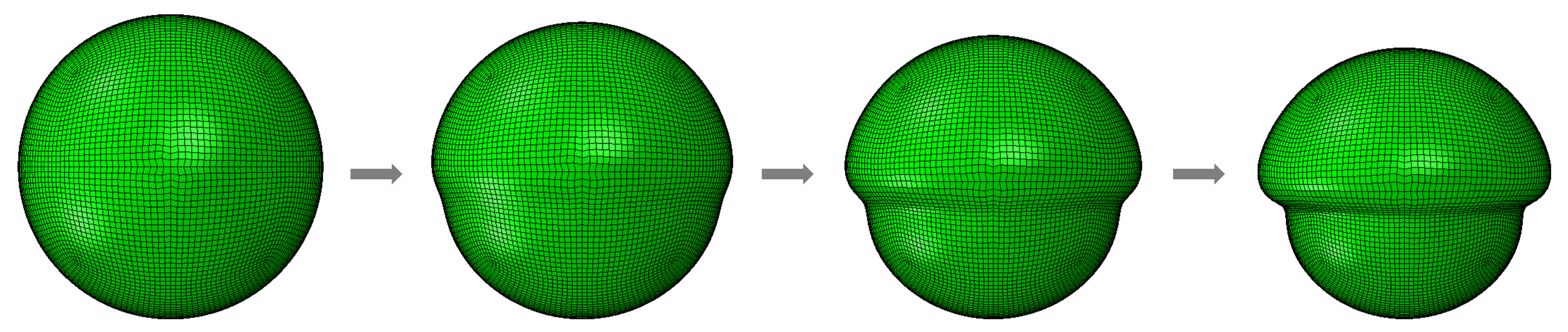}
	\caption{The invagination process of the cell sheet simulated by Abaqus software.}
	\label{fig:ev}
\end{figure}

The active stretches $\gamma_1$ and $\gamma_2$ drive the cell sheet transforms from the  initially spherical geometry into an invaginated configuration, characterized by an inward folding or indentation of the surface. By comparing $\gamma_1$ with $\gamma_2$, it becomes clear that the active stretch $\gamma_2$ at the equatorial region seems the key driver of the cell sheet's invagination.  This deformation pattern is a consequence of the spatially varying active stretches imposed on the cell sheet, mimicking the complex cellular processes that govern morphogenesis. The shape transformation is vividly depicted in Fig. \ref{fig:ev} using Abaqus simulations. 

\begin{figure}[h!]
	\centering
	\subfloat[]{\includegraphics[width=0.3\textwidth]{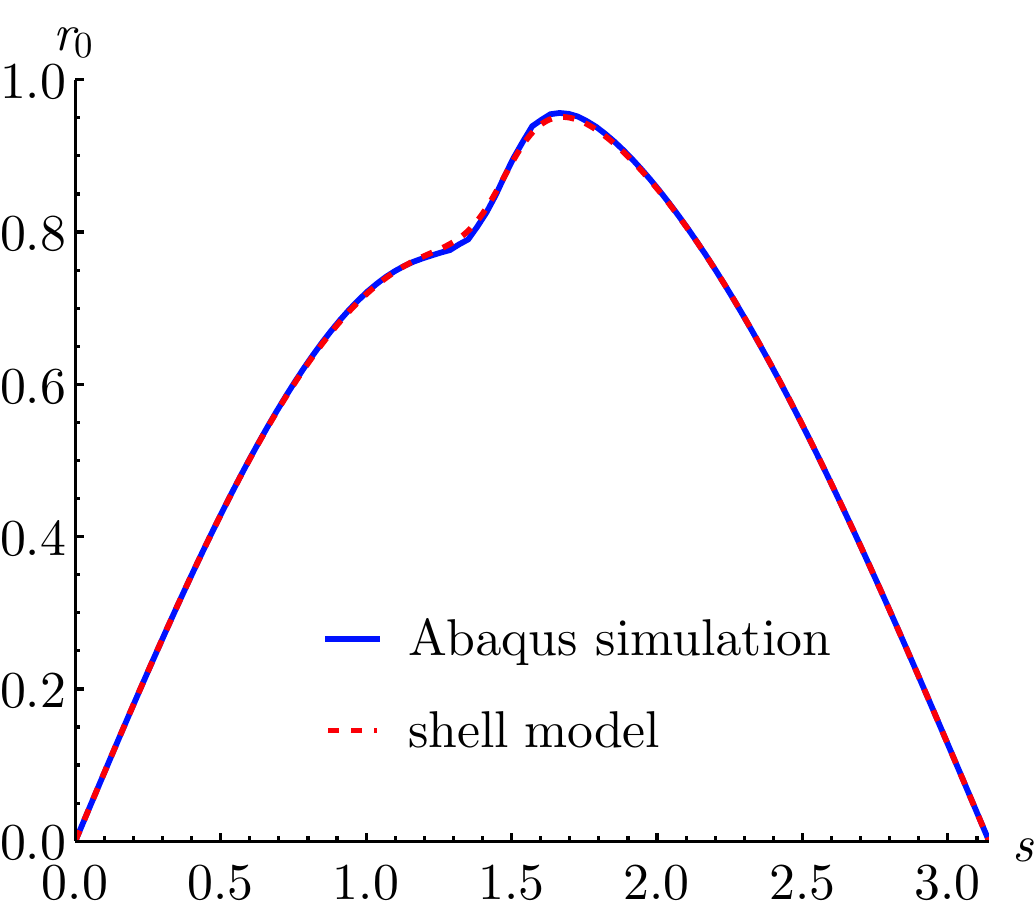}} \quad 
	\subfloat[]{\includegraphics[width=0.3\textwidth]{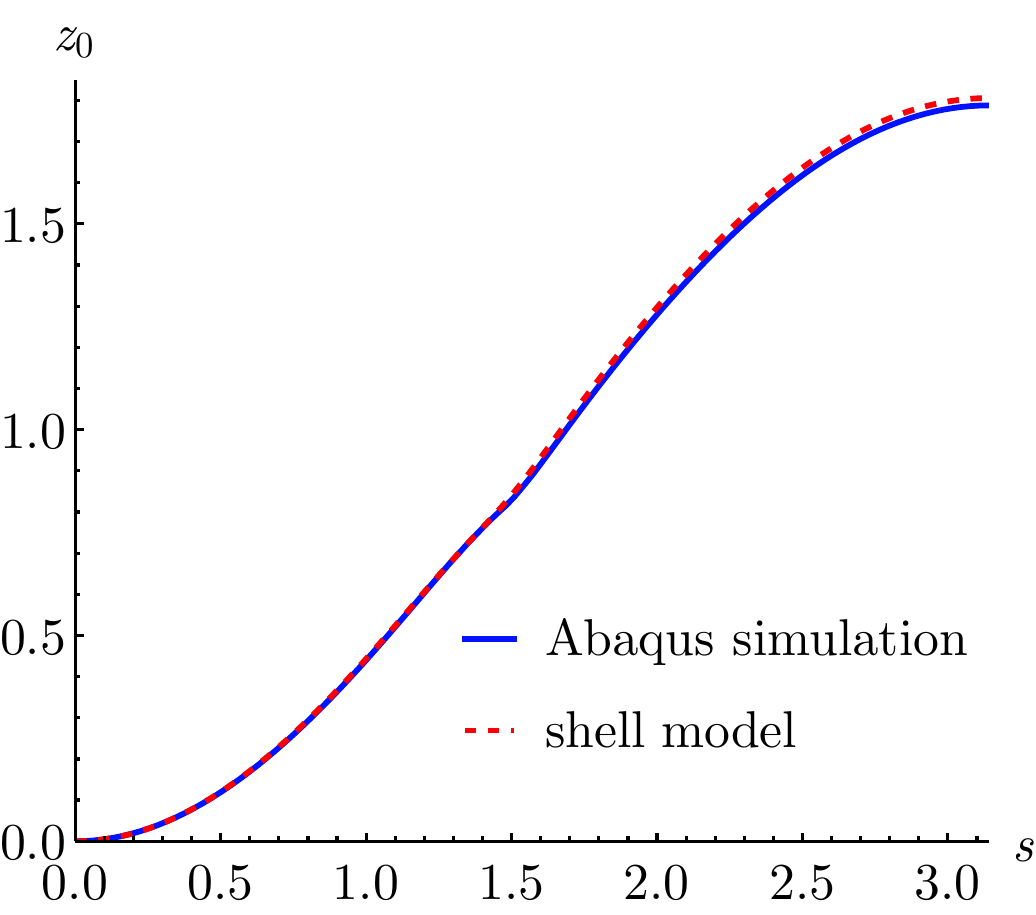}}\quad
	\subfloat[]{\includegraphics[width=0.28\textwidth]{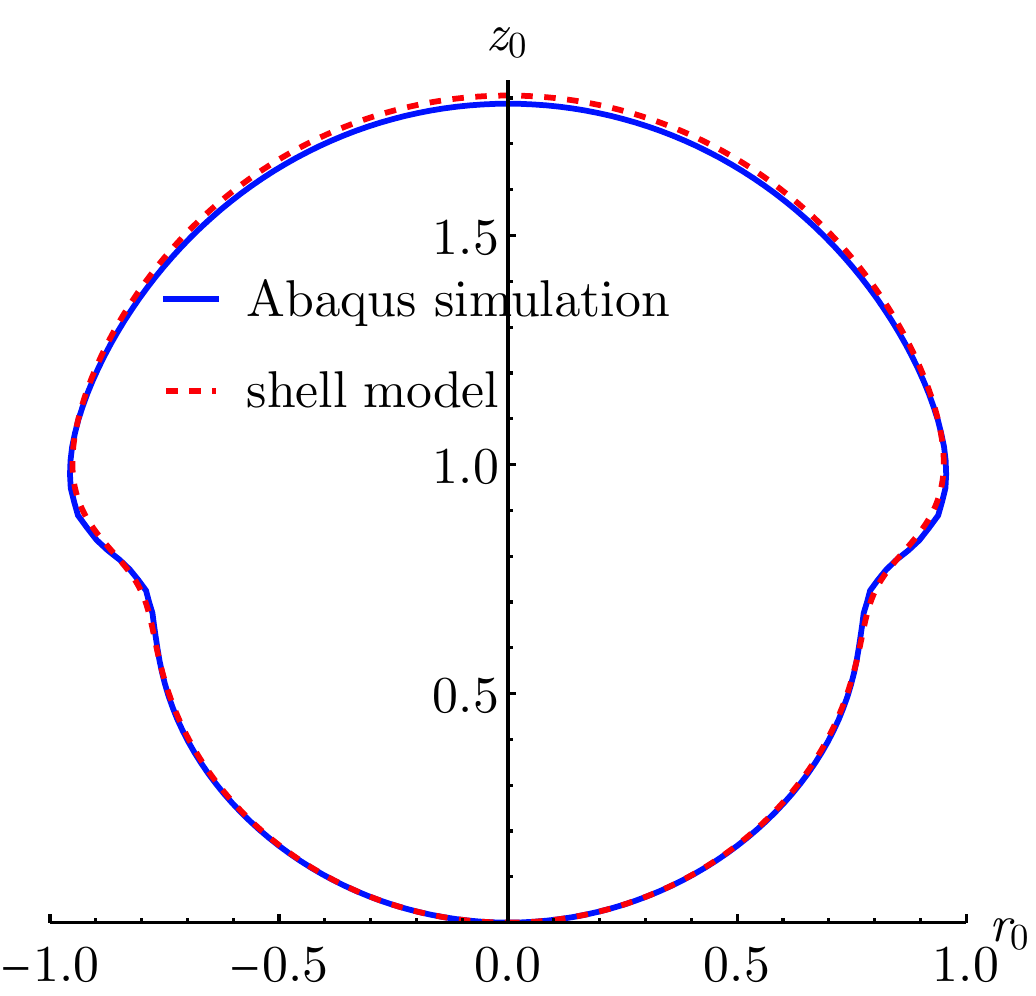}}
	\caption{The position of the actual middle surface given by Abaqus simulations and the shell model: (a,b) the radial and axial coordinates $r_0(s)$ and $z_0(s)$ and (c) the middle surface profile.
} 
	\label{fig:r0z0}
\end{figure}

\begin{figure}[h!]
	\centering
	\subfloat[]{\includegraphics[width=0.36\textwidth]{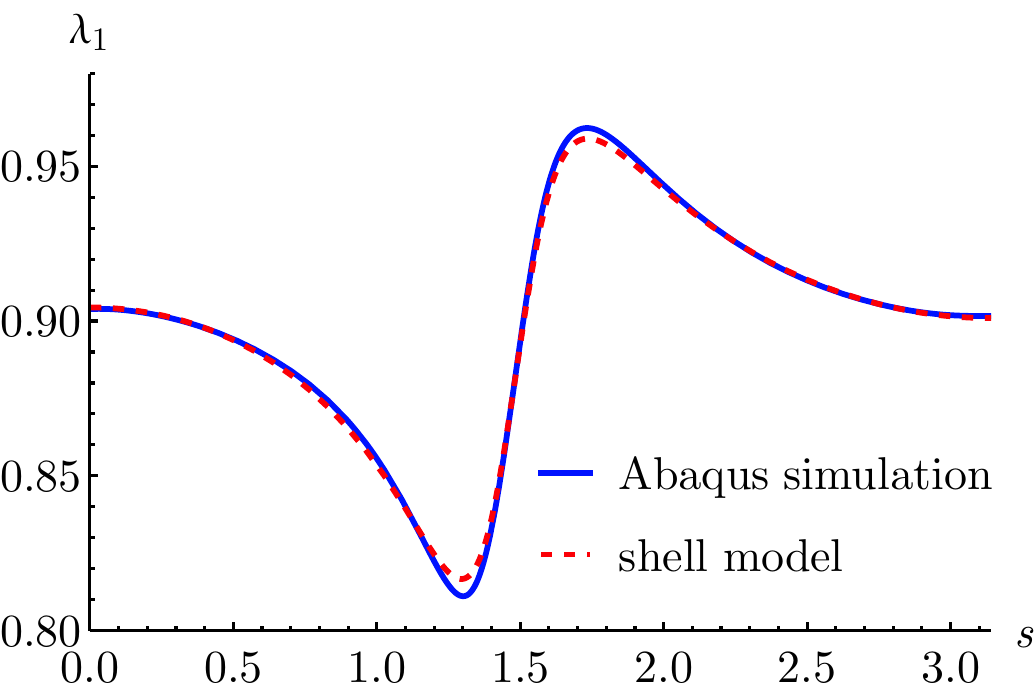}} \qquad \qquad \quad
	\subfloat[]{\includegraphics[width=0.36\textwidth]{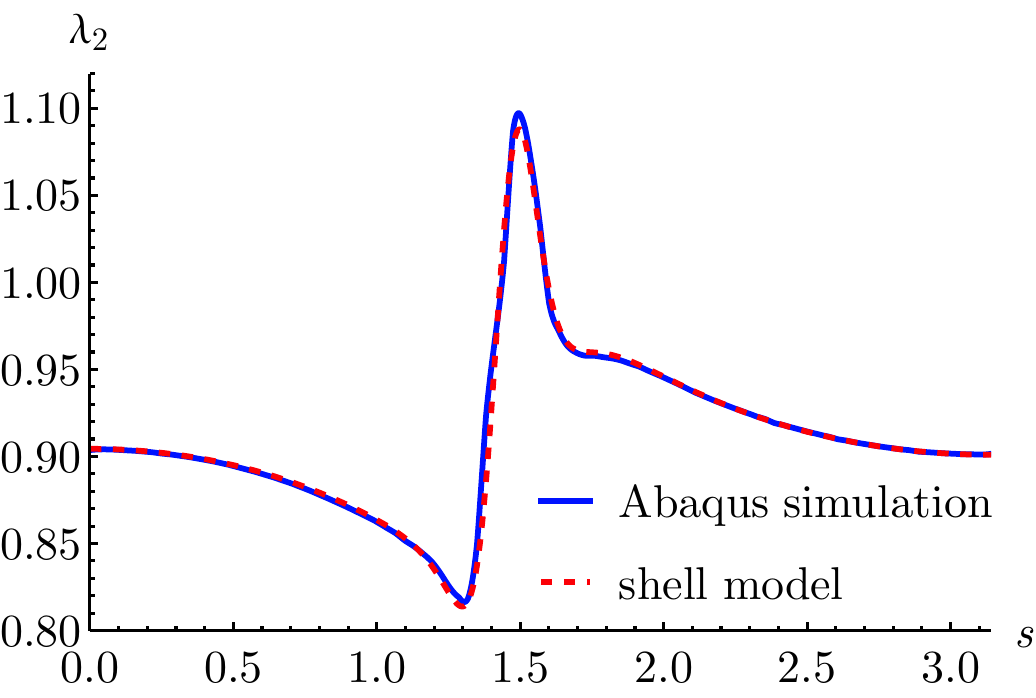} }
	\caption{The two principal stretches $\lambda_1$ and $\lambda_2$ of the middle surface  given by Abaqus simulations and the shell model.} 
	\label{fig:lambda}
\end{figure}

\begin{figure}[htbp!]
	\centering
	\subfloat[]{\includegraphics[scale=0.36]{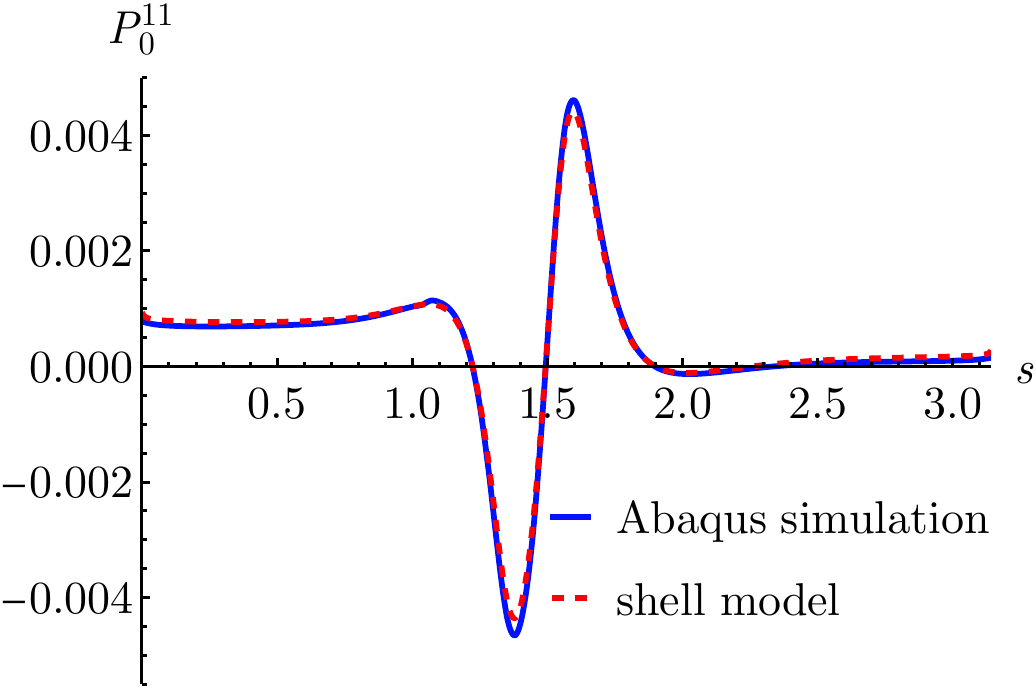}}\qquad \qquad \quad
	\subfloat[]{\includegraphics[scale=0.36]{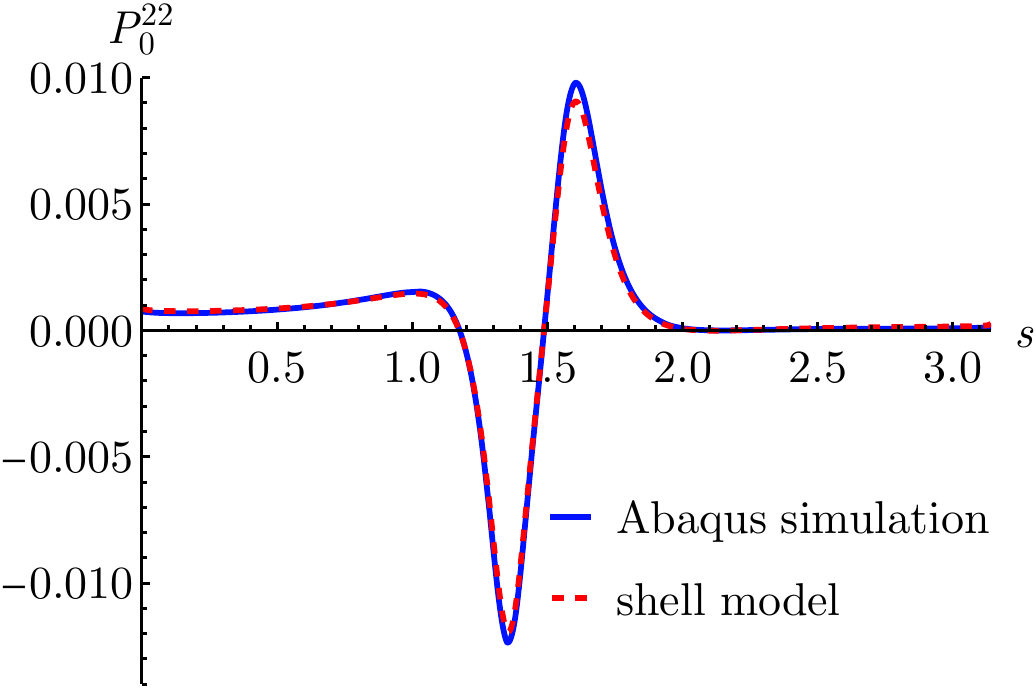}} 
	\caption{The leading-order  Piola-Kirchhoff stresses in the (a) azimuthal direction and (b)  meridional direction    given by Abaqus simulations and the shell model.}
	\label{fig:P0}
\end{figure}

To affirm the validity of our model, we have conducted a comprehensive comparison with 3D finite element simulations of Abaqus. The position coordinates $r_0(s)$ and $z_0(s)$ of the actual middle surface and its profile given by Abaqus simulations and our shell model are shown in Fig. \ref{fig:r0z0}. The associated principal stretches and first Piola-Kirchhoff  stress field are illustrated in Figures \ref{fig:lambda} and \ref{fig:P0}, respectively. It is seen that the results predicted by our shell model are in close agreement with Abaqus simulations, attesting to the model's accuracy and predictive strength in modeling the mechanics of biological tissues.

We want to discuss more about the actual stretches and the residual stresses. The two principal stretches of the middle surface in the azimuthal and meridional directions are given by 
\begin{align}
\lambda_1=\frac{r_0(s)}{\sin s},\quad \lambda_2=\sqrt{r'^2_0(s)+z'^2_0(s)},
\end{align}
which are shown in Fig. \ref{fig:lambda}. It is important to note that the magnitude of the actual stretches cannot be considered small, with a maximum value of up to $20\%$ near the equator at $s=2 \pi/5$. This suggests that a finite-strain shell theory may be a more appropriate choice for analysis. On the other hand, the existence of the residual stresses, as shown in Fig. \ref{fig:P0}, implies that the active stretches are incompatible since no external loadings are imposed. The good agreement between the model predictions and Abaqus simulations confirms that our shell model is suitable for analyzing incompatible growth.

\section{Conclusion}\label{sec:con}
We have derived a morphoelastic shell model to treat the finite-strain deformation of thin biological tissues, taking the compressibility into account. This model is based on  series expansions and the variational asymptotic method. By choosing appropriate base surface---the actual middle surface of the current configuration, we expand the  displacement relative to this base surface into series with respect to the thickness variable. Then by fixing the leading-order coefficient, the 3D energy functional is optimized to determine the higher-order coefficients, which leads to our optimal shell energy containing the leading-order coefficient as the only unknown. This morphoelastic shell energy retains the variational structure, and can be directly used for numerical simulation.  It incorporates both the stretching and bending effects, and is applicable to finite-strain deformations under various loading conditions. By a further step, optimizing the shell energy, we obtain the shell equations and boundary conditions. We verify that from the energy aspect, the morphoelastic shell model recovers the classical shell theories under appropriate scaling assumptions. And from the equation aspect, it is also asymptotically equivalent to Dai {\it et al.}'s shell model. Two 3D benchmark problems are used for validation and one biological example is studied for illustration of the present shel model.

It is important to acknowledge the underlying assumptions of our shell model. 
For the geometry of the shell, we have implicitly assumed  that the components of the curvature tensor are much smaller than the reciprocal of the thickness, i.e., $\kappa h \ll 1$ with $\kappa$ denoting a generic component of the curvature tensor.  In other words, the radius of curvature should be much larger than the thickness of the shell. 
This assumption is present since $\kappa   h$ appears in some expansions, such as in the expansion of $\mu(Z)$  in the energy (\ref{eq:EE}). This is a restriction of the shell geometry  for our shell model. Furthermore, for the surface loads, it is necessary that  $\bar{\bm{q}}$ should be no greater than $O(1)$ to maintain the equilibrium. This implies that the net traction should not exceed $O(h)$ in view of $\eqref{eq:qm}$, i.e., $\mu(h/2) \bm{q}^++\mu(-h/2) \bm{q}^- \leq O(h)$.  It is important to note that this condition still permits $\bm{q}^{-}$ and $\bm{q}^{+}$ to be of  $O(1)$.
In relation to the growth tensor,  since it is incorporated into the augmented  energy function, it is required to be integrable in the weak form of our theoretical framework, as shown in (\ref{eq:twod}), and to be first-order differentiable in the strong form, as articulated in (\ref{eq:pde}).


Despite these assumptions, our model does not impose \textit{ad hoc} constraints on strain or hyperelastic material types, thereby enhancing its broad applicability under complex loading conditions for a variety of hyperelastic materials.
 Moreover, our model's variational framework also streamlines its incorporation into numerical simulations. It provides a powerful tool to investigate the finite-strain deformations of thin biological shells. 
However, the model does have a limitation related to its ``strong form"---that is, while the derived shell equations and boundary conditions are mathematically sound and succinct, they might not offer the same degree of physical insight as the shell model introduced by Dai {\it et al.}
Conversely, while Dai {\it et al.}'s model provides a more intuitive physical interpretation, it does not maintain the variational principles when it comes to boundary conditions and tends to involve more  unknowns for a similar level of precision as our model.
In the future, we plan to further explore the numerical capabilities of our model.


A Mathematica code that produces the results presented in Section \ref{sec:example} is available on GitHub
(\url{https://github.com/xiangyudgut}).

\section*{Acknowledgments}
This work is supported the National Natural Science Foundation of China (Grant No. 12272055), and partially by Guangdong Provincial Key Laboratory 440 of Interdisciplinary Research and Application for Data Science, and
partially by Guangdong Provincial Key Laboratory
 of Interdisciplinary Research and Application for Data Science, BNU-HKBU United International College (Project No. 2022B1212010006), and Guangdong Basic and Applied Basic Research Foundation (Grant No. 2023A1515111141).
 
 The authors thank Dr. Zhanfeng Li of South China University of Technology for his help with the Abaqus simulations reported in this paper.

\appendix

\section{Replacement of  $\bm{y}_1$ by its membrane approximation}\label{app:x1}

This section justifies the claim that one can replace $\bm{y}_1$ by its membrane approximation $\bm{f}(\nabla\bm{r})$ in the energy functional \eqref{eq:inter} without influencing the latter's accuracy order. 

According to the perturbation method, we see from $\eqref{eq:mt}_1$ and \eqref{eq:mm} that $\bm{y}_1$  can be expanded as
\begin{align}\label{eq:A1}
\bm{y}_1=\bm{f}(\nabla\bm{r})+h^2  \bm{z}_1+O(h^3),
\end{align}
where $h^2 \bm{z}_1$ is the error term.
Inserting \eqref{eq:A1} into \eqref{eq:G} and expanding the resulting equation at $\bm{y}_1=\bm{f}(\nabla\bm{r})$, we obtain
\begin{align}\label{eq:GG}
\begin{split}
L= &\Big(h\bar{W}(\bm{F}_0)-h\bar{\bm{q}}\cdot\bm{r}-\bm{m}\cdot\bm{y}_1+h^3 L_3\Big)\Big|_{\bm{y}_1=\bm{f}(\nabla\bm{r})}\\
& +\underbrace{h^3[\bar{W}_{\bm{F}}(\nabla\bm{r}+\bm{f}(\nabla\bm{r})\otimes\bm{n})\bm{n}-\bm{m}]\cdot \bm{z}_1}_{=0}+O(h^4).
\end{split}
\end{align}
The error term (the last second term) vanishes thanks to \eqref{eq:mm}, proving the claim announced.

\section{Derivation of shell equations and boundary conditions}\label{app:B}
In this section, we derive the shell equations and boundary conditions from the 
variation of the two-dimensional shell energy.

In view of \eqref{eq:G1}, we see that the variation of shell energy \eqref{eq:twod} is given by
\begin{align}\label{eq:vtwod}
\begin{split}
\delta\mathcal{E}_\text{2d}=&\int_{\Omega} (\bm{T}:\nabla \delta \bm{r}+\mathcal{M}\therefore\nabla\nabla\delta\bm{r}-h\bar{\bm{q}}\cdot\delta\bm{r})\,dA-\int_{\partial\Omega_t}(\bm{p}_0\cdot \delta\bm{r}+\bm{p}_1\cdot \mathcal{F}:\nabla \delta \bm{r})\,dS,
\end{split}
\end{align}
where $\bm{T}:=\partial L/\partial\nabla\bm{r}$, $\mathcal{M}$ and $\mathcal{F}$ are defined in \eqref{eq:NMN}, and $\therefore$ means the triple contraction of two tensors (i.e., $(\mathcal{A}\therefore \mathcal{B})_{i_1\dots i_{p-3}j_4\dots j_q}:=\mathcal{A}_{i_1\dots i_{p-3}klm}\mathcal{B}_{klmj_4\dots j_q}$). The product rule for derivatives implies 
\begin{align}\label{eq:MM}
\mathcal{M}\therefore\nabla\nabla\delta\bm{r}=\nabla\cdot (\mathcal{M}^{T_{312}}:\nabla\nabla\bm{r})-(\nabla\cdot \mathcal{M}^{T_{312}}):\nabla \nabla\bm{r},
\end{align}
where $(\mathcal{M}^{T_{312}})_{ijk}=\mathcal{M}_{kij}$ denotes the transpose of $\mathcal{M}$ with respect to the permutation $123\to 312$. Substituting \eqref{eq:MM} into \eqref{eq:vtwod} and applying the surface divergence theorem, we obtain
\begin{align}
\begin{split}
\delta\mathcal{E}_\text{2d}=&\int_{\Omega} (\bm{N}:\nabla \delta \bm{r}-h\bar{\bm{q}}\cdot\delta\bm{r})\,dA+\int_{\partial\Omega_t}\left[(\mathcal{M}\cdot\bm{\nu}-\mathcal{F}^{T_{231}}\cdot \bm{p}_1):\nabla\delta\bm{r}-\bm{p}_0\cdot \delta\bm{r}\right]\,dS\\
&+\int_{\partial\Omega_u} (\mathcal{M}\cdot\bm{\nu}):\nabla\delta\bm{r}\,dS,
\end{split}
\end{align}
where $\bm{N}=\bm{T}-\nabla\cdot \mathcal{M}^{T_{312}}$ and $(\mathcal{F}^{T_{231}})_{ijk}=\mathcal{F}_{jki}$. 

A repeated application of the product rule and surface divergence theorem yields
\begin{align}\label{eq:v2d}
\begin{split}
\delta\mathcal{E}_\text{2d}=&-\int_{\Omega} (\nabla\cdot\bm{N}^T+h\bar{\bm{q}})\cdot\delta\bm{r}\,dA+\int_{\partial\Omega_t}\left[(\bm{N}\bm{\nu}-\bm{p}_0)\cdot\delta\bm{r}+(\mathcal{M}\cdot\bm{\nu}-\mathcal{F}^{T_{231}}\cdot \bm{p}_1):\nabla\delta\bm{r}\right]\,dS\\
&+\int_{\partial\Omega_u} (\bm{N}\bm{\nu}\cdot \delta\bm{r}+(\mathcal{M}\cdot\bm{\nu}):\nabla\delta\bm{r})\,dS.
\end{split}
\end{align}
On the boundary $\partial\Omega_t$, we have the decomposition $\nabla\delta\bm{r}=\delta\bm{r}_{,S}\otimes\bm{\tau}+\delta\bm{r}_{,\nu}\otimes\bm{\nu}$. Upon using integration by parts, we can simplify \eqref{eq:v2d} as
\begin{align}
\begin{split}
\delta\mathcal{E}_\text{2d}=&-\int_{\Omega} (\nabla\cdot\bm{N}^T+h\bar{\bm{q}})\cdot\bm\delta\bm{r}\,dA+\int_{\partial\Omega_t} \left[\bm{N}\bm{\nu}-\bm{p}_0-(\mathcal{M}[\bm{\tau},\bm{\nu}])_{,S}+(\mathcal{F}^{T_{231}}[\bm{\tau},\bm{p}_1])_{,S}\right]\cdot\delta\bm{r} \,dS\\
&+\int_{\partial\Omega_t} \left(\mathcal{M}[\bm{\nu},\bm{\nu}]-\mathcal{F}^{T_{231}}[\bm{\nu},\bm{p}_1]\right)\cdot\delta\bm{r}_{,\nu} \,dS+\int_{\partial\Omega_u} [\bm{N}\bm{\nu}-(\mathcal{M}[\bm{\tau},\bm{\nu}])_{,S}]\cdot\delta\bm{r}\,dS\\
&+\int_{\partial\Omega_u} \mathcal{M}[\bm{\nu},\bm{\nu}]\cdot\delta\bm{r}_{,\nu} \,dS,
\end{split}
\end{align}
where $\mathcal{M}[\bm{\tau},\bm{\nu}]:=(\mathcal{M}\cdot\bm{\nu})\cdot\bm{\tau}$, etc. Setting $\delta\mathcal{E}_\text{2d}=0$, one recovers the shell  equations and boundary conditions given in \eqref{eq:pde}--\eqref{eq:u}.

\section{Simplification of the bending strain tensor in linear elasticity}\label{app:bending}

In this section, we derive the bending strain tensor explicitly in terms of the displacements of the middle surface, in the context of linear elasticity.	

The bending strain tensor is defined in \eqref{Eq_bending_strain} as
\begin{equation}
\label{Eq_bending}
\bm{\rho}=\nabla\bm{u}_0\bm{\kappa}+\nabla\bm{u}_1.
\end{equation}
Let us denoted by $\bm{\alpha}$ the vector $\bm{\alpha}=\bm{n}\cdot \nabla \bm{u}_0$. Noting that  $\bm{I}=\bm{1}+\bm{n}\otimes\bm{n}$, we have the decomposition
\begin{align}\label{eq:C3_1}
\nabla\bm{u}_0=\bm{1}\nabla\bm{u}_0+\bm{n}\otimes\bm{\alpha},\quad \nabla\bm{u}_1=\bm{1} \nabla \bm{u}_1+\bm{n}\otimes (\bm{n}\cdot \nabla\bm{u}_1).
\end{align}
From $\eqref{eq:C3_1}_1$, we have
\begin{align}\label{eq:C3}
\nabla\bm{u}_0\bm{\kappa}=\bm{1}\nabla\bm{u}_0\bm{\kappa}+\bm{n}\otimes\bm{\kappa}\bm{\alpha}.
\end{align}
And from $\eqref{eq:u1u2}_1$, $\bm{u}_1$ is expressed as
\begin{align}
\bm{u}_1=-\bm{\alpha}-\eta\tr(\bm{1}\nabla\bm{u}_0)\bm{n}.
\end{align}
Thus the surface gradient of $\bm{u}_1$ is given by
\begin{align}\label{eq:du1}
\nabla\bm{u}_1=-\nabla\bm{\alpha}-\eta \nabla[\tr(\bm{1}\nabla\bm{u}_0)\bm{n}].
\end{align}
It follows from \eqref{eq:du1} that 
\begin{equation}
\begin{aligned}
\label{eq:C6}
\bm{n}\cdot \nabla\bm{u}_1&=-\bm{n}\cdot\nabla\bm{\alpha}-\eta\bm{n}\cdot \nabla[\tr(\bm{1}\nabla\bm{u}_0)\bm{n}]\\
&=-\bm{\kappa}\bm{\alpha}-\eta \nabla \tr(\bm{1}\nabla\bm{u}_0),\\
\bm{1} \nabla \bm{u}_1&=\bm{1}\{-\nabla \bm{\alpha} -\eta \nabla [\tr(\bm{1}\nabla\bm{u}_0)\bm{n}]\}\\
&=-\bm{1}[\nabla(\bm{n}\cdot\nabla\bm{u}_0)]-\eta \bm{1}\nabla [\tr(\bm{1}\nabla\bm{u}_0)\bm{n}]\\
&=-\bm{1}(\bm{n}\cdot\nabla\nabla\bm{u}_0)+(\nabla\bm{u}_0)^T\bm{\kappa}+\eta \tr(\bm{1}\nabla\bm{u}_0)\bm{\kappa}.
\end{aligned}
\end{equation}
where we have used the fact that $\bm{n}$ and $\bm{\alpha}$ are orthogonal and $\bm{\kappa}=-\nabla\bm{n}$. 

Substituting \eqref{eq:C6} back into $\eqref{eq:du1}_2$ which is further taken into \eqref{Eq_bending} with  \eqref{eq:C3}, 
 we obtain the following explicit expression for the bending strain tensor 
\begin{align}
\bm{\rho}=\nabla\bm{u}_0\bm{\kappa}+\nabla\bm{u}_1=-\bm{1}(\bm{n}\cdot\nabla\nabla\bm{u}_0)+(2\bm{\varepsilon}+\eta \tr(\bm{\varepsilon})\bm{1})\bm{\kappa}-\eta\bm{n}\otimes \nabla \tr(\bm{\varepsilon}),
\end{align}  
where $\bm{\varepsilon}=\frac{1}{2}(\bm{1}\nabla\bm{u}_0+(\bm{1}\nabla\bm{u}_0)^T)$ is the in-plane stretching strain and we have used the equality $(\bm{1}\nabla\bm{u}_0)^T\bm{\kappa}=(\nabla\bm{u}_0)^T\bm{\kappa}$ since $\bm{\kappa}$ is an in-plane tensor.


\begin{thebibliography}{53}
	\expandafter\ifx\csname natexlab\endcsname\relax\def\natexlab#1{#1}\fi
	\providecommand{\url}[1]{\texttt{#1}}
	\providecommand{\href}[2]{#2}
	\providecommand{\path}[1]{#1}
	\providecommand{\DOIprefix}{doi:}
	\providecommand{\ArXivprefix}{arXiv:}
	\providecommand{\URLprefix}{URL: }
	\providecommand{\Pubmedprefix}{pmid:}
	\providecommand{\doi}[1]{\href{http://dx.doi.org/#1}{\path{#1}}}
	\providecommand{\Pubmed}[1]{\href{pmid:#1}{\path{#1}}}
	\providecommand{\bibinfo}[2]{#2}
	\ifx\xfnm\relax \def\xfnm[#1]{\unskip,\space#1}\fi
	\bibitem[{Audoly \& Hutchinson(2016)}]{audoly2016analysis}
	\bibinfo{author}{Audoly, B.}, \& \bibinfo{author}{Hutchinson, J.~W.}
	(\bibinfo{year}{2016}).
	\newblock \bibinfo{title}{Analysis of necking based on a one-dimensional
		model}.
	\newblock {\it \bibinfo{journal}{Journal of the Mechanics and Physics of
			Solids}\/},  {\it \bibinfo{volume}{97}\/}, \bibinfo{pages}{68--91}.
	\bibitem[{Audoly \& Lestringant(2021)}]{audoly2021asymptotic}
	\bibinfo{author}{Audoly, B.}, \& \bibinfo{author}{Lestringant, C.}
	(\bibinfo{year}{2021}).
	\newblock \bibinfo{title}{Asymptotic derivation of high-order rod models from
		non-linear 3d elasticity}.
	\newblock {\it \bibinfo{journal}{Journal of the Mechanics and Physics of
			Solids}\/},  {\it \bibinfo{volume}{148}\/}, \bibinfo{pages}{104264}.
	\bibitem[{Berdichevskii(1979)}]{berdichevskii1979variational}
	\bibinfo{author}{Berdichevskii, V.} (\bibinfo{year}{1979}).
	\newblock \bibinfo{title}{Variational-asymptotic method of constructing a
		theory of shells: Pmm vol. 43, no. 4, 1979, pp. 664--687}.
	\newblock {\it \bibinfo{journal}{Journal of Applied Mathematics and
			Mechanics}\/},  {\it \bibinfo{volume}{43}\/}, \bibinfo{pages}{711--736}.
	\bibitem[{Birzle et~al.(2018)Birzle, Martin, Yoshihara, Uhlig \&
		Wall}]{birzle2018experimental}
	\bibinfo{author}{Birzle, A.~M.}, \bibinfo{author}{Martin, C.},
	\bibinfo{author}{Yoshihara, L.}, \bibinfo{author}{Uhlig, S.}, \&
	\bibinfo{author}{Wall, W.~A.} (\bibinfo{year}{2018}).
	\newblock \bibinfo{title}{Experimental characterization and model
		identification of the nonlinear compressible material behavior of lung
		parenchyma}.
	\newblock {\it \bibinfo{journal}{Journal of the Mechanical Behavior of
			Biomedical Materials}\/},  {\it \bibinfo{volume}{77}\/},
	\bibinfo{pages}{754--763}.
	\bibitem[{Carrera \& Zozulya(2022)}]{carrera2022carrera}
	\bibinfo{author}{Carrera, E.}, \& \bibinfo{author}{Zozulya, V.}
	(\bibinfo{year}{2022}).
	\newblock \bibinfo{title}{Carrera unified formulation (cuf) for the micropolar
		plates and shells. i. higher order theory}.
	\newblock {\it \bibinfo{journal}{Mechanics of Advanced Materials and
			Structures}\/},  {\it \bibinfo{volume}{29}\/}, \bibinfo{pages}{773--795}.
	\bibitem[{Chadwick(1999)}]{chadwick1999continuum}
	\bibinfo{author}{Chadwick, P.} (\bibinfo{year}{1999}).
	\newblock {\it \bibinfo{title}{Continuum mechanics: concise theory and
			problems}\/}.
	\newblock \bibinfo{publisher}{Courier Corporation}.
	\bibitem[{Chen et~al.(2021)Chen, Ciarletta \& Dai}]{chen2021physical}
	\bibinfo{author}{Chen, X.}, \bibinfo{author}{Ciarletta, P.}, \&
	\bibinfo{author}{Dai, H.-H.} (\bibinfo{year}{2021}).
	\newblock \bibinfo{title}{Physical principles of morphogenesis in mushrooms}.
	\newblock {\it \bibinfo{journal}{Physical Review E}\/},  {\it
		\bibinfo{volume}{103}\/}, \bibinfo{pages}{022412}.
	\bibitem[{Chen \& Dai(2020)}]{chen2020stress}
	\bibinfo{author}{Chen, X.}, \& \bibinfo{author}{Dai, H.-H.}
	(\bibinfo{year}{2020}).
	\newblock \bibinfo{title}{Stress-free configurations induced by a family of
		locally incompatible growth functions}.
	\newblock {\it \bibinfo{journal}{Journal of the Mechanics and Physics of
			Solids}\/},  {\it \bibinfo{volume}{137}\/}, \bibinfo{pages}{103834}.
	\bibitem[{Chen et~al.(2022)Chen, Shen, Li, Gu \& Wang}]{chen2022generating}
	\bibinfo{author}{Chen, X.}, \bibinfo{author}{Shen, Y.}, \bibinfo{author}{Li,
		Z.}, \bibinfo{author}{Gu, D.}, \& \bibinfo{author}{Wang, J.}
	(\bibinfo{year}{2022}).
	\newblock \bibinfo{title}{Generating complex fold patterns through stress-free
		deformation induced by growth}.
	\newblock {\it \bibinfo{journal}{Journal of the Mechanics and Physics of
			Solids}\/},  {\it \bibinfo{volume}{159}\/}, \bibinfo{pages}{104702}.
	\bibitem[{Chesler et~al.(2004)Chesler, Thompson-Figueroa~and \&
		Millburne}]{chesler2004measurements}
	\bibinfo{author}{Chesler, N.~C.}, \bibinfo{author}{Thompson-Figueroa~and, J.},
	\& \bibinfo{author}{Millburne, K.} (\bibinfo{year}{2004}).
	\newblock \bibinfo{title}{Measurements of mouse pulmonary artery biomechanics}.
	\newblock {\it \bibinfo{journal}{Journal of Biomechanical Engineering}\/},
	{\it \bibinfo{volume}{126}\/}, \bibinfo{pages}{309--313}.
	\bibitem[{Chuong \& Fung(1984)}]{chuong1984compressibility}
	\bibinfo{author}{Chuong, C.}, \& \bibinfo{author}{Fung, Y.}
	(\bibinfo{year}{1984}).
	\newblock \bibinfo{title}{Compressibility and constitutive equation of arterial
		wall in radial compression experiments}.
	\newblock {\it \bibinfo{journal}{Journal of Biomechanics}\/},  {\it
		\bibinfo{volume}{17}\/}, \bibinfo{pages}{35--40}.
	\bibitem[{Ciarlet(2005)}]{ciarlet2005introduction}
	\bibinfo{author}{Ciarlet, P.~G.} (\bibinfo{year}{2005}).
	\newblock \bibinfo{title}{An introduction to differential geometry with
		applications to elasticity}.
	\newblock {\it \bibinfo{journal}{Journal of Elasticity}\/},  {\it
		\bibinfo{volume}{78}\/}, \bibinfo{pages}{1--215}.
	\bibitem[{Ciarletta \& Amar(2012)}]{ciarletta2012growth}
	\bibinfo{author}{Ciarletta, P.}, \& \bibinfo{author}{Amar, M.~B.}
	(\bibinfo{year}{2012}).
	\newblock \bibinfo{title}{Growth instabilities and folding in tubular organs: a
		variational method in non-linear elasticity}.
	\newblock {\it \bibinfo{journal}{International Journal of Non-Linear
			Mechanics}\/},  {\it \bibinfo{volume}{47}\/}, \bibinfo{pages}{248--257}.
	\bibitem[{Dai \& Ben~Amar(2022)}]{dai2022minimizing}
	\bibinfo{author}{Dai, A.}, \& \bibinfo{author}{Ben~Amar, M.}
	(\bibinfo{year}{2022}).
	\newblock \bibinfo{title}{Minimizing the elastic energy of growing leaves by
		conformal mapping}.
	\newblock {\it \bibinfo{journal}{Physical Review Letters}\/},  {\it
		\bibinfo{volume}{129}\/}, \bibinfo{pages}{218101}.
	\bibitem[{Dai \& Song(2014)}]{dai2014consistent}
	\bibinfo{author}{Dai, H.-H.}, \& \bibinfo{author}{Song, Z.}
	(\bibinfo{year}{2014}).
	\newblock \bibinfo{title}{On a consistent finite-strain plate theory based on
		three-dimensional energy principle}.
	\newblock {\it \bibinfo{journal}{Proceedings of the Royal Society A:
			Mathematical, Physical and Engineering Sciences}\/},  {\it
		\bibinfo{volume}{470}\/}, \bibinfo{pages}{20140494}.
	\bibitem[{Dervaux \& Ben~Amar(2008)}]{dervaux2008morphogenesis}
	\bibinfo{author}{Dervaux, J.}, \& \bibinfo{author}{Ben~Amar, M.}
	(\bibinfo{year}{2008}).
	\newblock \bibinfo{title}{Morphogenesis of growing soft tissues}.
	\newblock {\it \bibinfo{journal}{Physical Review Letters}\/},  {\it
		\bibinfo{volume}{101}\/}, \bibinfo{pages}{068101}.
	\bibitem[{Dervaux et~al.(2009)Dervaux, Ciarletta \&
		Ben~Amar}]{dervaux2009morphogenesis}
	\bibinfo{author}{Dervaux, J.}, \bibinfo{author}{Ciarletta, P.}, \&
	\bibinfo{author}{Ben~Amar, M.} (\bibinfo{year}{2009}).
	\newblock \bibinfo{title}{Morphogenesis of thin hyperelastic plates: a
		constitutive theory of biological growth in the f{\"o}ppl--von k{\'a}rm{\'a}n
		limit}.
	\newblock {\it \bibinfo{journal}{Journal of the Mechanics and Physics of
			Solids}\/},  {\it \bibinfo{volume}{57}\/}, \bibinfo{pages}{458--471}.
	\bibitem[{Dhatt et~al.(2012)Dhatt, Lefran{\c{c}}ois \&
		Touzot}]{dhatt2012finite}
	\bibinfo{author}{Dhatt, G.}, \bibinfo{author}{Lefran{\c{c}}ois, E.}, \&
	\bibinfo{author}{Touzot, G.} (\bibinfo{year}{2012}).
	\newblock {\it \bibinfo{title}{Finite element method}\/}.
	\newblock \bibinfo{publisher}{John Wiley \& Sons}.
	\bibitem[{Di~Puccio et~al.(2012)Di~Puccio, Celi \& Forte}]{di2012review}
	\bibinfo{author}{Di~Puccio, F.}, \bibinfo{author}{Celi, S.}, \&
	\bibinfo{author}{Forte, P.} (\bibinfo{year}{2012}).
	\newblock \bibinfo{title}{Review of experimental investigations on
		compressibility of arteries and introduction of a new apparatus}.
	\newblock {\it \bibinfo{journal}{Experimental Mechanics}\/},  {\it
		\bibinfo{volume}{52}\/}, \bibinfo{pages}{895--902}.
	\bibitem[{Do~Carmo(2016)}]{do2016differential}
	\bibinfo{author}{Do~Carmo, M.~P.} (\bibinfo{year}{2016}).
	\newblock {\it \bibinfo{title}{Differential geometry of curves and surfaces:
			revised and updated second edition}\/}.
	\newblock \bibinfo{publisher}{Courier Dover Publications}.
	\bibitem[{Du et~al.(2020)Du, Dai, Wang \& Wang}]{du2020analytical}
	\bibinfo{author}{Du, P.}, \bibinfo{author}{Dai, H.-H.}, \bibinfo{author}{Wang,
		J.}, \& \bibinfo{author}{Wang, Q.} (\bibinfo{year}{2020}).
	\newblock \bibinfo{title}{Analytical study on growth-induced bending
		deformations of multi-layered hyperelastic plates}.
	\newblock {\it \bibinfo{journal}{International Journal of Non-Linear
			Mechanics}\/},  {\it \bibinfo{volume}{119}\/}, \bibinfo{pages}{103370}.
	\bibitem[{Du et~al.(2023{\natexlab{a}})Du, Li, Chen \& Wang}]{du2023general}
	\bibinfo{author}{Du, P.}, \bibinfo{author}{Li, Z.}, \bibinfo{author}{Chen, X.},
	\& \bibinfo{author}{Wang, J.} (\bibinfo{year}{2023}{\natexlab{a}}).
	\newblock \bibinfo{title}{A general multi-layered hyperelastic plate theory for
		growth-induced deformations in soft material samples}.
	\newblock {\it \bibinfo{journal}{Applied Mathematical Modelling}\/},  {\it
		\bibinfo{volume}{115}\/}, \bibinfo{pages}{300--336}.
	\bibitem[{Du et~al.(2023{\natexlab{b}})Du, Wang \& Wang}]{du2023simplified}
	\bibinfo{author}{Du, P.}, \bibinfo{author}{Wang, F.-F.}, \&
	\bibinfo{author}{Wang, J.} (\bibinfo{year}{2023}{\natexlab{b}}).
	\newblock \bibinfo{title}{On a simplified multi-layered plate model of growth:
		Asymptotic analyses and numerical implementation}.
	\newblock {\it \bibinfo{journal}{Thin-Walled Structures}\/},  {\it
		\bibinfo{volume}{191}\/}, \bibinfo{pages}{111100}.
	\bibitem[{Friesecke et~al.(2006)Friesecke, James \&
		M{\"u}ller}]{friesecke2006hierarchy}
	\bibinfo{author}{Friesecke, G.}, \bibinfo{author}{James, R.~D.}, \&
	\bibinfo{author}{M{\"u}ller, S.} (\bibinfo{year}{2006}).
	\newblock \bibinfo{title}{A hierarchy of plate models derived from nonlinear
		elasticity by gamma-convergence}.
	\newblock {\it \bibinfo{journal}{Archive for rational mechanics and
			analysis}\/},  {\it \bibinfo{volume}{180}\/}, \bibinfo{pages}{183--236}.
	\bibitem[{Guo et~al.(2015)Guo, Dai, Han, Xie, Chao \& Chen}]{guo2015fast}
	\bibinfo{author}{Guo, Q.}, \bibinfo{author}{Dai, E.}, \bibinfo{author}{Han,
		X.}, \bibinfo{author}{Xie, S.}, \bibinfo{author}{Chao, E.}, \&
	\bibinfo{author}{Chen, Z.} (\bibinfo{year}{2015}).
	\newblock \bibinfo{title}{Fast nastic motion of plants and bioinspired
		structures}.
	\newblock {\it \bibinfo{journal}{Journal of the Royal Society Interface}\/},
	{\it \bibinfo{volume}{12}\/}, \bibinfo{pages}{20150598}.
	\bibitem[{Haas \& Goldstein(2021)}]{haas2021morphoelasticity}
	\bibinfo{author}{Haas, P.~A.}, \& \bibinfo{author}{Goldstein, R.~E.}
	(\bibinfo{year}{2021}).
	\newblock \bibinfo{title}{Morphoelasticity of large bending deformations of
		cell sheets during development}.
	\newblock {\it \bibinfo{journal}{Physical Review E}\/},  {\it
		\bibinfo{volume}{103}\/}, \bibinfo{pages}{022411}.
	\bibitem[{H{\"o}hn \& Hallmann(2011)}]{hohn2011there}
	\bibinfo{author}{H{\"o}hn, S.}, \& \bibinfo{author}{Hallmann, A.}
	(\bibinfo{year}{2011}).
	\newblock \bibinfo{title}{There is more than one way to turn a spherical
		cellular monolayer inside out: type b embryo inversion in volvox globator}.
	\newblock {\it \bibinfo{journal}{BMC Biology}\/},  {\it \bibinfo{volume}{9}\/},
	\bibinfo{pages}{1--26}.
	\bibitem[{Keller et~al.(2003)Keller, Davidson \& Shook}]{keller2003we}
	\bibinfo{author}{Keller, R.}, \bibinfo{author}{Davidson, L.~A.}, \&
	\bibinfo{author}{Shook, D.~R.} (\bibinfo{year}{2003}).
	\newblock \bibinfo{title}{How we are shaped: the biomechanics of gastrulation}.
	\newblock {\it \bibinfo{journal}{Differentiation: original article}\/},  {\it
		\bibinfo{volume}{71}\/}, \bibinfo{pages}{171--205}.
	\bibitem[{Lestringant \& Audoly(2020)}]{lestringant2020asymptotically}
	\bibinfo{author}{Lestringant, C.}, \& \bibinfo{author}{Audoly, B.}
	(\bibinfo{year}{2020}).
	\newblock \bibinfo{title}{Asymptotically exact strain-gradient models for
		nonlinear slender elastic structures: a systematic derivation method}.
	\newblock {\it \bibinfo{journal}{Journal of the Mechanics and Physics of
			Solids}\/},  {\it \bibinfo{volume}{136}\/}, \bibinfo{pages}{103730}.
	\bibitem[{Li et~al.(2023)Li, Wang, Hossain \& Kadapa}]{li2023general}
	\bibinfo{author}{Li, Z.}, \bibinfo{author}{Wang, J.}, \bibinfo{author}{Hossain,
		M.}, \& \bibinfo{author}{Kadapa, C.} (\bibinfo{year}{2023}).
	\newblock \bibinfo{title}{A general theoretical scheme for shape-programming of
		incompressible hyperelastic shells through differential growth}.
	\newblock {\it \bibinfo{journal}{International Journal of Solids and
			Structures}\/},  {\it \bibinfo{volume}{265}\/}, \bibinfo{pages}{112128}.
	\bibitem[{Li et~al.(2022)Li, Wang, Du, Kadapa, Hossain \&
		Wang}]{li2022analytical}
	\bibinfo{author}{Li, Z.}, \bibinfo{author}{Wang, Q.}, \bibinfo{author}{Du, P.},
	\bibinfo{author}{Kadapa, C.}, \bibinfo{author}{Hossain, M.}, \&
	\bibinfo{author}{Wang, J.} (\bibinfo{year}{2022}).
	\newblock \bibinfo{title}{Analytical study on growth-induced axisymmetric
		deformations and shape-control of circular hyperelastic plates}.
	\newblock {\it \bibinfo{journal}{International Journal of Engineering
			Science}\/},  {\it \bibinfo{volume}{170}\/}, \bibinfo{pages}{103594}.
	\bibitem[{Mehta et~al.(2022)Mehta, Raju \& Saxena}]{mehta2022wrinkling}
	\bibinfo{author}{Mehta, S.}, \bibinfo{author}{Raju, G.}, \&
	\bibinfo{author}{Saxena, P.} (\bibinfo{year}{2022}).
	\newblock \bibinfo{title}{Wrinkling as a mechanical instability in growing
		annular hyperelastic plates}.
	\newblock {\it \bibinfo{journal}{International Journal of Mechanical
			Sciences}\/},  {\it \bibinfo{volume}{229}\/}, \bibinfo{pages}{107481}.
	\bibitem[{Nolan \& McGarry(2016)}]{nolan2016compressibility}
	\bibinfo{author}{Nolan, D.}, \& \bibinfo{author}{McGarry, J.}
	(\bibinfo{year}{2016}).
	\newblock \bibinfo{title}{On the compressibility of arterial tissue}.
	\newblock {\it \bibinfo{journal}{Annals of Biomedical Engineering}\/},  {\it
		\bibinfo{volume}{44}\/}, \bibinfo{pages}{993--1007}.
	\bibitem[{Ogden(1984)}]{ogden1984non}
	\bibinfo{author}{Ogden, R.~W.} (\bibinfo{year}{1984}).
	\newblock {\it \bibinfo{title}{Non-linear elastic deformations}\/}.
	\newblock \bibinfo{publisher}{Ellis Horwood, New York}.
	\bibitem[{Pan(2014)}]{pan2014tensor}
	\bibinfo{author}{Pan, R.} (\bibinfo{year}{2014}).
	\newblock \bibinfo{title}{Tensor transpose and its properties}.
	\newblock {\it \bibinfo{journal}{arXiv preprint arXiv:1411.1503}\/}, .
	\bibitem[{Rausch \& Kuhl(2014)}]{rausch2014mechanics}
	\bibinfo{author}{Rausch, M.~K.}, \& \bibinfo{author}{Kuhl, E.}
	(\bibinfo{year}{2014}).
	\newblock \bibinfo{title}{On the mechanics of growing thin biological
		membranes}.
	\newblock {\it \bibinfo{journal}{Journal of the Mechanics and Physics of
			Solids}\/},  {\it \bibinfo{volume}{63}\/}, \bibinfo{pages}{128--140}.
	\bibitem[{Reddy(2006)}]{reddy2006theory}
	\bibinfo{author}{Reddy, J.~N.} (\bibinfo{year}{2006}).
	\newblock {\it \bibinfo{title}{Theory and analysis of elastic plates and
			shells}\/}.
	\newblock \bibinfo{publisher}{CRC press}.
	\bibitem[{Rodriguez et~al.(1994)Rodriguez, Hoger \&
		McCulloch}]{rodriguez1994stress}
	\bibinfo{author}{Rodriguez, E.~K.}, \bibinfo{author}{Hoger, A.}, \&
	\bibinfo{author}{McCulloch, A.~D.} (\bibinfo{year}{1994}).
	\newblock \bibinfo{title}{Stress-dependent finite growth in soft elastic
		tissues}.
	\newblock {\it \bibinfo{journal}{Journal of Biomechanics}\/},  {\it
		\bibinfo{volume}{27}\/}, \bibinfo{pages}{455--467}.
	\bibitem[{Rudraraju et~al.(2019)Rudraraju, Moulton, Chirat, Goriely \&
		Garikipati}]{rudraraju2019computational}
	\bibinfo{author}{Rudraraju, S.}, \bibinfo{author}{Moulton, D.~E.},
	\bibinfo{author}{Chirat, R.}, \bibinfo{author}{Goriely, A.}, \&
	\bibinfo{author}{Garikipati, K.} (\bibinfo{year}{2019}).
	\newblock \bibinfo{title}{A computational framework for the morpho-elastic
		development of molluskan shells by surface and volume growth}.
	\newblock {\it \bibinfo{journal}{PLoS Computational Biology}\/},  {\it
		\bibinfo{volume}{15}\/}, \bibinfo{pages}{e1007213}.
	\bibitem[{Song \& Dai(2016)}]{song2016consistent}
	\bibinfo{author}{Song, Z.}, \& \bibinfo{author}{Dai, H.-H.}
	(\bibinfo{year}{2016}).
	\newblock \bibinfo{title}{On a consistent finite-strain shell theory based on
		3-d nonlinear elasticity}.
	\newblock {\it \bibinfo{journal}{International Journal of Solids and
			Structures}\/},  {\it \bibinfo{volume}{97}\/}, \bibinfo{pages}{137--149}.
	\bibitem[{Steigmann et~al.(2023)Steigmann, Birsan \&
		Shirani}]{steigmann2023lecture}
	\bibinfo{author}{Steigmann, D.}, \bibinfo{author}{Birsan, M.}, \&
	\bibinfo{author}{Shirani} (\bibinfo{year}{2023}).
	\newblock {\it \bibinfo{title}{Lecture notes on the theory of plates and
			shells. Classical and modern developments}\/}.
	\newblock \bibinfo{publisher}{Cham: Springer}.
	\bibitem[{Steigmann(2008)}]{steigmann2008two}
	\bibinfo{author}{Steigmann, D.~J.} (\bibinfo{year}{2008}).
	\newblock \bibinfo{title}{Two-dimensional models for the combined bending and
		stretching of plates and shells based on three-dimensional linear
		elasticity}.
	\newblock {\it \bibinfo{journal}{International Journal of Engineering
			Science}\/},  {\it \bibinfo{volume}{46}\/}, \bibinfo{pages}{654--676}.
	\bibitem[{Steigmann(2012)}]{steigmann2012extension}
	\bibinfo{author}{Steigmann, D.~J.} (\bibinfo{year}{2012}).
	\newblock \bibinfo{title}{Extension of koiter’s linear shell theory to
		materials exhibiting arbitrary symmetry}.
	\newblock {\it \bibinfo{journal}{International Journal of Engineering
			Science}\/},  {\it \bibinfo{volume}{51}\/}, \bibinfo{pages}{216--232}.
	\bibitem[{Steigmann(2015)}]{steigmann2015mechanics}
	\bibinfo{author}{Steigmann, D.~J.} (\bibinfo{year}{2015}).
	\newblock \bibinfo{title}{Mechanics of materially uniform thin films}.
	\newblock {\it \bibinfo{journal}{Mathematics and Mechanics of Solids}\/},  {\it
		\bibinfo{volume}{20}\/}, \bibinfo{pages}{309--326}.
	\bibitem[{Steigmann \& Ogden(1999)}]{steigmann1999elastic}
	\bibinfo{author}{Steigmann, D.~J.}, \& \bibinfo{author}{Ogden, R.}
	(\bibinfo{year}{1999}).
	\newblock \bibinfo{title}{Elastic surface—substrate interactions}.
	\newblock {\it \bibinfo{journal}{Proceedings of the Royal Society of London.
			Series A: Mathematical, Physical and Engineering Sciences}\/},  {\it
		\bibinfo{volume}{455}\/}, \bibinfo{pages}{437--474}.
	\bibitem[{Tickner \& Sacks(1967)}]{tickner1967theory}
	\bibinfo{author}{Tickner, E.~G.}, \& \bibinfo{author}{Sacks, A.~H.}
	(\bibinfo{year}{1967}).
	\newblock \bibinfo{title}{A theory for the static elastic behavior of blood
		vessels}.
	\newblock {\it \bibinfo{journal}{Biorheology}\/},  {\it \bibinfo{volume}{4}\/},
	\bibinfo{pages}{151--168}.
	\bibitem[{Ventsel \& Krauthammer(2002)}]{ventsel2002thin}
	\bibinfo{author}{Ventsel, E.}, \& \bibinfo{author}{Krauthammer, T.}
	(\bibinfo{year}{2002}).
	\newblock \bibinfo{title}{Thin plates and shells: theory, analysis, and
		applications}.
	\bibitem[{Wang et~al.(2018)Wang, Steigmann, Wang \& Dai}]{wang2018consistent}
	\bibinfo{author}{Wang, J.}, \bibinfo{author}{Steigmann, D.},
	\bibinfo{author}{Wang, F.-F.}, \& \bibinfo{author}{Dai, H.-H.}
	(\bibinfo{year}{2018}).
	\newblock \bibinfo{title}{On a consistent finite-strain plate theory of
		growth}.
	\newblock {\it \bibinfo{journal}{Journal of the Mechanics and Physics of
			Solids}\/},  {\it \bibinfo{volume}{111}\/}, \bibinfo{pages}{184--214}.
	\bibitem[{{Wolfram Research Inc}(2024)}]{wolfram2024mathematica}
	\bibinfo{author}{{Wolfram Research Inc}} (\bibinfo{year}{2024}).
	\newblock {\it \bibinfo{title}{Mathematica, Version 14.0, URL
			https://www.wolfram.com/mathematica}\/}.
	\newblock \bibinfo{publisher}{champaign, IL}.
	\bibitem[{Xu et~al.(2020)Xu, Fu \& Yang}]{xu2020water}
	\bibinfo{author}{Xu, F.}, \bibinfo{author}{Fu, C.}, \& \bibinfo{author}{Yang,
		Y.} (\bibinfo{year}{2020}).
	\newblock \bibinfo{title}{Water affects morphogenesis of growing aquatic plant
		leaves}.
	\newblock {\it \bibinfo{journal}{Physical Review Letters}\/},  {\it
		\bibinfo{volume}{124}\/}, \bibinfo{pages}{038003}.
	\bibitem[{Yin et~al.(2022)Yin, Li \& Feng}]{yin2022three}
	\bibinfo{author}{Yin, S.}, \bibinfo{author}{Li, B.}, \& \bibinfo{author}{Feng,
		X.-Q.} (\bibinfo{year}{2022}).
	\newblock \bibinfo{title}{Three-dimensional chiral morphodynamics of
		chemomechanical active shells}.
	\newblock {\it \bibinfo{journal}{Proceedings of the National Academy of
			Sciences}\/},  {\it \bibinfo{volume}{119}\/}, \bibinfo{pages}{e2206159119}.
	\bibitem[{Yossef et~al.(2017)Yossef, Farajian, Gilad, Willenz, Gutman \&
		Yosibash}]{yossef2017further}
	\bibinfo{author}{Yossef, O.~E.}, \bibinfo{author}{Farajian, M.},
	\bibinfo{author}{Gilad, I.}, \bibinfo{author}{Willenz, U.},
	\bibinfo{author}{Gutman, N.}, \& \bibinfo{author}{Yosibash, Z.}
	(\bibinfo{year}{2017}).
	\newblock \bibinfo{title}{Further experimental evidence of the compressibility
		of arteries}.
	\newblock {\it \bibinfo{journal}{Journal of the Mechanical Behavior of
			Biomedical Materials}\/},  {\it \bibinfo{volume}{65}\/},
	\bibinfo{pages}{177--189}.
	\bibitem[{Yu et~al.(2020)Yu, Fu \& Dai}]{yu2020refined}
	\bibinfo{author}{Yu, X.}, \bibinfo{author}{Fu, Y.}, \& \bibinfo{author}{Dai,
		H.-H.} (\bibinfo{year}{2020}).
	\newblock \bibinfo{title}{A refined dynamic finite-strain shell theory for
		incompressible hyperelastic materials: equations and two-dimensional shell
		virtual work principle}.
	\newblock {\it \bibinfo{journal}{Proceedings of the Royal Society A}\/},  {\it
		\bibinfo{volume}{476}\/}, \bibinfo{pages}{20200031}.
	
\end{thebibliography}

\end{document}